\documentstyle[11pt,aaspp4,flushrt]{article}

\begin{document}

\newcommand{\spp}{\mbox {S$^{++}$}}
\newcommand{\op}{\mbox {O$^+$}}
\newcommand{\opp}{\mbox {O$^{++}$}}
\newcommand{\teff}{\mbox {$T_{\!\!\em eff}$}}
\newcommand{\tstar}{\mbox {$T_*$}}
\newcommand{\qh}{\mbox {$Q_{H^0}$}}
\newcommand{\x}{\mbox {$R_{23}$}}
\newcommand{\stt}{\mbox {$S_{23}$}}
\newcommand{\etap}{$\eta^\prime$}
\newcommand{\ariii}{\mbox {[Ar III]\thinspace $\lambda$7135}}
\newcommand{\sii}{\mbox {[S II]\thinspace $\lambda\lambda$6716,\thinspace 6731}}
\newcommand{\oii}{\mbox {[O II]\thinspace $\lambda$3727}}
\newcommand{\siii}{\mbox {[S III]\thinspace $\lambda\lambda$9069,\thinspace 9532}}
\newcommand{\oiii}{\mbox {[O III]\thinspace $\lambda\lambda$4959,\thinspace 5007}}
\newcommand{\nii}{\mbox {[N II]\thinspace $\lambda\lambda$6548,\thinspace 6583}}
\newcommand{\neiii}{\mbox {[Ne III]\thinspace $\lambda$3869}}
\newcommand{\msun}{\mbox {${\rm M_\odot}$}}
\newcommand{\zsun}{\mbox {${\rm Z_{\odot}}$}}
\newcommand{\lsun}{\mbox {${\rm L_\odot}$}}
\newcommand{\angs}{\mbox{~\AA}}
\newcommand{\lines}{\mbox {$\lambda\lambda$}}
\newcommand{\hii}{H\thinspace II}
\newcommand{\ew}{EW(H$\beta$)}
\newcommand{\halpha}{H$\alpha$} 
\newcommand{\ha}{\mbox {H$\alpha$}}
\newcommand{\hbeta}{H$\beta$}
\newcommand{\hgamma}{H$\gamma$}
\newcommand{\hb}{\mbox {H$\beta$}}
\newcommand{\cf}{cf.\/\ } 
\newcommand{\eg}{e.g.\/}
\newcommand{\ie}{i.e.\/\  } 
\newcommand{\etal}{et\thinspace al.\ }   


\lefthead{Kennicutt, Bresolin, French & Martin} 
\righthead{HII region diagnostics}

\title{An Empirical Test and Calibration of \hii\/ Region Diagnostics}

\author{Robert C. Kennicutt, Jr.,\altaffilmark{1,2}\ 
Fabio Bresolin,\altaffilmark{3,4}\ 
Howard French,\altaffilmark{5,6}\ 
Pierre Martin\altaffilmark{7}
}
\altaffiltext{1}{Steward Observatory, 
University of Arizona, Tucson, AZ 85721}

\altaffiltext{2}{Visiting Astronomer, Cerro Tololo Interamerican 
Observatory, National Optical Astronomical Observatories, which are
operated by AURA, Inc., under contract with the National Science 
Foundation.}

\altaffiltext{3}{European Southern Observatory, 
Karl-Schwarzschild-Str. 2, D-85748 Garching b. M\"{u}nchen, Germany}

\altaffiltext{4}{Present address: Institut fur Physik und Astronomie, 
Scheinerstr. 1, D-81679, Munich, Germany}

\altaffiltext{5}{Department of Astronomy, University of 
Minnesota, 116 Church Street SE, Minneapolis, MN 55455}

\altaffiltext{6}{Honeywell Technology Center, 3660 Technology Drive,
Minneapolis, MN 55418}

\altaffiltext{7}{CFHT, P.O. Box 1597, Kamuela, HI 96743}

\begin{abstract} 

We present spectrophotometry in the 3600--9700 \AA\ region for a
sample of 39 \hii\ regions in the Galaxy and Magellanic Clouds,
for which independent information is available on the spectral
types and effective temperatures of the ionizing stars.  The
spectra have been used to evaluate nebular diagnostics
of stellar temperature, metal abundance, and ionization parameter,
and compare the observed behavior of the line indices with predictions
of nebular photoionization models.  We observe a strong degeneracy
between forbidden-line sequences produced by changes in stellar \teff\/
and metal abundance, which severely complicates the application of
many forbidden-line diagnostics to extragalactic \hii\ regions.  Our
data confirm however that the Edmunds \& Pagel [\ion{O}{2}]$+$[\ion{O}{3}]
abundance index and the V\'{\i}lchez \& Pagel \etap\ index provide
more robust diagnostics
of metal abundance and stellar effective temperature, respectively.
A comparison of the fractional helium ionization of the \hii\ regions
with stellar temperature confirms the reliability of the spectral
type vs \teff\/ calibration for the relevant temperature range
\teff\ $\le$ 38000 K.  We use empirical relations between the nebular
hardness indices and \teff\/ to reinvestigate the case for systematic
variations in the stellar effective temperatures and the upper IMFs of 
massive stars in extragalactic \hii\ regions.  The data are consistent
with a significant softening of the ionizing spectra (consistent with
cooler stellar temperatures) with increasing metal abundance, especially
for $Z \le$ \zsun.  However unresolved degeneracies between $Z$ and \teff\ 
still complicate the interpretation of this result.

\end{abstract}


\section{INTRODUCTION}

Over the past 15 years high-quality emission-line spectra have been obtained
for hundreds of \hii\ regions in nearby spiral and irregular galaxies.  The
primary use of these data has been to study the chemical composition
patterns in galactic disks (e.g., McCall, Rybski, \& Shields 1985, 
Vila-Costas \& Edmunds 1992, Zaritsky, Kennicutt, \& Huchra 1994, Dinerstein 
1996).  However with the availability of state-of-the-art photoionization codes
such as {\sc cloudy} (Ferland \etal 1998), combined with stellar population 
synthesis codes for young clusters (e.g., Leitherer \& Heckman 1995,
Leitherer et al. 1999), \hii\ region spectra are being applied increasingly
to measure the ages and initial mass functions (IMFs) of the ionizing
star clusters (e.g., V\'{\i}lchez \& Pagel 1988, Shields 1990, Stasi\'nska
\& Leitherer 1996, Bresolin, Kennicutt, \& Garnett 1999, hereafter BKG).  
Similar techniques have been applied to model the infrared emission-line 
spectra of IR-luminous starbursts, and the results have been interpreted as 
evidence for an anomalous IMF in these objects (e.g., Rieke et al. 1993).

A basic limitation in this approach is its reliance on a long chain of 
theoretical inputs: stellar evolution models as functions of stellar
mass, chemical composition and mass-loss rates; stellar atmosphere models
as functions of effective temperature, surface gravity, chemical composition,
and mass-loss properties, and photoionization models for the surrounding
\hii\ region.  The derived nebular abundances tend to be insensitive 
to the details of these models, but the inferred properties of the ionizing
stars are sensitive to the model inputs at every step in the chain.
The problem is especially acute when using \hii\ region or starburst 
spectra to constrain
the effective temperatures and IMFs of the ionizing stars (e.g., Mathis 1985,
McCall et al.~1985, V\'{\i}lchez \& Pagel 1988, BKG).
Most studies show evidence for a systematic softening of the
ionizing continuum of \hii\ regions with increasing metal abundance, but
the absolute range in stellar effective temperatures is highly model 
dependent, and the case for a systematic variation in the upper 
IMF is shaky at best (BKG and references therein).

Another approach, which can circumvent many of these difficulties,
is to test the nebular diagnostics empirically, by obtaining spectra
for nearby \hii\ regions which are ionized by stars of known spectral type. 
This makes it possible to anchor the nebular indicators of stellar
temperature directly to the stellar spectral classification system.
Furthermore, by modeling these \hii\ regions with the same methods that are 
applied to extragalactic regions one can directly assess the reliability
of the model-based approach.

This empirical approach was first explored by Chopinet \&
Lortet-Zuckermann (1976) and Kaler (1978) to calibrate the 
[\ion{O}{3}]$\lambda$5007/\hbeta\ ratio as an indicator of stellar
effective temperature.  For studies of extragalactic \hii\ regions 
and starbursts a more robust hardness index than [\ion{O}{3}]/\hbeta\ 
is desirable, because the excitation of [\ion{O}{3}] varies locally within
\hii\ regions, and it is sensitive to other physical parameters 
(e.g., metal abundance, ionization parameter, dust) which can vary
systematically in galaxies (e.g., Kennicutt 1984, Shields 1990).
Other hardness indices have been developed for extragalactic applications
(e.g., V\'{\i}lchez \& Pagel 1988), but their reliability has not been tested 
empirically. 

In this paper we report the results of a spectrophotometric study 
of 39 \hii\ regions in the Galaxy, the LMC, and the SMC, which is aimed at 
providing an empirical foundation for nebular-based measurements of the stellar
ionizing continuum and IMF in galaxies.  We use the spectra to evaluate
several nebular diagnostics of massive stellar populations, and to
compare the known properties of the stars in these regions with
those inferred from photoionization modeling.
Our study is mainly motivated by applications to stellar 
temperatures and IMFs, but the approach is more generally applicable to  
spectral modeling of extragalactic \hii\ regions and starbursts.
This paper is complementary in approach to a recent analysis of Oey
et al. (2000), which has addressed many of these same questions using
detailed point-by-point observations of 4 \hii\ regions in the LMC.

The remainder of this paper is organized as follows.  
The \hii\/ region sample and the observations are
described in \S~2 and \S~3, respectively.  In \S~4 we analyze 
the behavior of the principal diagnostic line ratios and investigate
several commonly applied abundance and spectral hardness indicators.
In \S~5 we apply the empirical calibration to constrain the range of
stellar effective temperature and upper IMF in galactic disks.

\section{\hii\/ REGION SAMPLE AND PROPERTIES}

Since the first prerequisite for this work is knowledge of the exciting
stars in the calibrating \hii\ regions, the ideal approach is to 
observe small Galactic \hii\ regions ionized by single stars (or a few stars)
of known spectral type.  Such ``calibrating" objects comprise approximately
half of our sample, while the remaining objects,
ionized by larger OB associations or clusters, provide the link to
the brighter class of \hii\ regions observed in external galaxies.

The calibrating \hii\ regions were selected to span the maximum available 
range in stellar spectral type and effective temperature.  
Preference was given to nebulae with 
angular diameters that fit within the length of the spectrograph slit
(3\arcmin\ $-$ 6\arcmin), and to regions with low to moderate extinction,
with $E(B-V) < 2$ mag in most cases.
Most of our sample was drawn from the northern survey  
of small Galactic \hii\ regions defined by Hunter \& Massey (1990; =HM90),
and the southern sample of \hii\ regions studied by Shaver et al.~(1983)
in their abundance survey of the Milky Way.
Identifications and spectral types for the ionizing stars were taken
directly from HM90 for the northern sample, and from a variety of published
sources for the southern objects.  Most of the latter 
derive from classification work in the 1970's by Georgelin and 
collaborators (e.g., Georgelin, Georgelin, \& Roux 1973, Chopinet,
Georgelin, \& Lortet-Zuckermann 1973).
Table 1 lists the main properties of the \hii\ regions in our sample,
including the sources of the stellar spectral types and oxygen abundances.
Column 8 indicates the observatory used to obtain the spectra, and
asterisks denote objects in the calibrating subsample.

It is relatively easy to compile an adequate sample of small \hii\ regions
ionized by single stars of type O6V and later (\teff\ $\le$ 42000).
However this becomes nearly impossible for hotter O stars, because
such stars almost always form in rich OB associations along with   
numerous later-type stars.  The contribution of the cooler
stars to the ionizing continuum can be significant, and this composite 
ionization needs to be taken into account in the determination of \teff.
We observed several of these large Galactic \hii\/ regions,  
including M8, M16, M17, the Rosette nebula (S275), the Carina nebula,
and the giant \hii\ region NGC~3603.  We used the most recent studies of
their stellar contents, including near-infrared observations in many cases,
to identify all of the principal ionizing stars.  
In order to extend our coverage to even higher stellar 
temperatures, we also observed four \hii\ regions ionized by early-type 
Wolf-Rayet stars, the Galactic regions RCW~5 = NGC~2359 (WN4), RCW~48 = 
NGC~3199 (WN5), and the LMC regions DEM~174 (WN4p) and DEM~231 (WN3).  

The remainder of our sample consists of much larger \hii\ regions
located in the LMC and SMC.  Most of these objects are physically
distinct from the smaller Galactic \hii\ regions, being ionized by
much more populous OB associations, and as such are less suitable for 
our empirical calibration of nebular diagnostics.  However these
objects are directly analagous to the more distant extragalactic \hii\ regions,
and their stellar contents have been cataloged by various authors
(Table 1), so we can analyze them on the same basis as the
smaller Galactic \hii\ regions.

To assign a stellar effective temperature to the \hii\/ regions we have
adopted the spectral type--\teff\/ calibration by Vacca, Garmany and
Shull (1996), together with the stellar ionizing fluxes predicted by
Schaerer \& de Koter (1997).  The resulting temperatures and luminosities
are listed in columns 4 and 5 of Table 1.  In the case of an ensemble of 
hot stars, a mean temperature was derived by summing the He$^0$ and 
H$^0$ ionizing fluxes of all of the stars, and calculating the 
effective temperature of a class V 
star having the same $Q_1/Q_0$ ratio (He$^0$ to H$^0$ total ionizing
flux ratio).  Effective temperatures for the W-R stars were estimated
using the semi-empirical calibration of Esteban et al. (1993).
More detailed information on the stellar contents
of the individual \hii\ regions is given in the Appendix.

Oxygen abundances, expressed as 12+$\log$(O/H), are given in column 6, and
their sources are indicated in column 7.  When abundances were 
unavailable in the literature and could not be measured directly from
our spectra, we estimated the abundance using the empirical indicator
\x=([O~III]\lines4959,5007+[O~II]\lines3726,3729)/\hbeta\/  (Pagel \etal
1979), as calibrated by Edmunds and Pagel (1984).  These estimates are
listed in parentheses in Table 1.  The accuracy of the \x-based abundances
is of order $\pm$0.2 dex, but this is sufficient for our purposes.

Figure~1 shows the range of stellar temperatures and oxygen abundances
for the \hii\ regions in our sample.   Squares and round points denote 
calibrating and non-calibrating \hii\ regions with
direct (electron temperature based) abundance measurements, respectively.
Open triangles indicate those regions with empirical (\x) abundance
estimates.  The \hii\ regions cover a stellar temperature range of
32000 $-$ 48000 K, with the Wolf-Rayet nebulae extending to 67000 K.
The 28 Galactic \hii\/ regions span a range of O/H abundances of
$-0.5 \le [O/H] \le 0.0$ (assuming 
12+$\log$(O/H)$_\odot$ = 8.9), while the 9 LMC regions cluster around
[O/H] $\simeq$ $-$0.4, and the single SMC region (N66 = NGC~346) 
has [O/H] = $-$0.7.  

The range of abundances in the sample is important for the interpretation
which follows, because changes in metal abundance can mimic many of the 
trends in nebular spectra that are produced by changes in stellar 
temperature or IMF.  For lack of a better prescription we have adopted a 
single conversion of spectral type to \teff\/ in our analysis, independent
of abundance, and this conceivably could introduce some coupling between
the two parameters (\S~4.3).  Fortunately, Fig.~1 shows that $Z/Z_\odot$ 
and \teff\ are largely decoupled from each other in our sample, and this
should allow us to separate their effects on the nebular spectra.

\section{OBSERVATIONS AND DATA REDUCTION}

Our data consist of spectrophotometry covering the wavelength range 
3600--9700 \AA.  The extended coverage to
the near-infrared includes the \siii\/ doublet,
which when combined with measurements of \sii, \oii, and \oiii\/
provides a robust measure of the hardness 
of the ionizing stellar continuum (Mathis 1985, V\'{\i}lchez \& Pagel 1988).

\subsection{Steward Observatory Data}

Spectra of the northern sample of Galactic \hii\/ regions were
obtained in 1993 Oct and 1995 Jun with the B\&C CCD spectrograph
on the Bok telescope, equipped 
with a thinned Loral 800$\times$1200 element CCD detector.  Two grating
settings were used to cover the full spectral range.  
A 400 g~mm$^{-1}$ grating blazed at 4800 \AA\ provided coverage of the 
3600--6900~\AA\ region, hereafter denoted as the ``blue"
spectra.  A second set of ``red" spectra were obtained 
with a 400 g~mm$^{-1}$ grating blazed at 7500 \AA\ to cover the
6400--9700 \AA\ region.  
A slit width of 2\farcs5 provided a spectral resolution 
of 8 \AA\ FWHM.  The full spatial coverage of a spectrum 
was 3\farcm5, with a single pixel on the detector projecting 
to 0\farcs83 on the sky.  Spectra of NGC~7538 and NGC~7635 were obtained in
in 1988 Nov using the same spectrograph but with a 800$\times$800 element 
Texas Instruments CCD detector.  The same setup was used for the red spectra,
but the blue spectra were obtained with a 600 g~mm$^{-1}$ grating
blazed at 3570 \AA, to cover the region between 3700--5100 \AA.

Figure~2 shows images of the \hii\ regions, with the slit
positions superimposed.  Most of the images were taken from the Digitized
Sky Surveys\footnotemark.  The \hii\ regions were observed at a fixed position  
extending E--W  through the
center of the nebula (but offset from bright stars to avoid
contamination of the spectrum), and also in a drift scanned mode, where
the image of the \hii\/ region was trailed across the slit to provide
an integrated spectrum.  Comparison of the fixed and scanned spectra 
(integrated over the length of the slit in both instances) reveals
surprisingly small differences in the main diagnostic line ratios, 
indicating that the fixed pointings provide a representative sampling
of the nebula.  This is illustrated in Figure~3, 
which compares the behavior of [\ion{O}{3}]/\hbeta\ vs [\ion{N}{2}]/\halpha\
and [\ion{S}{2}]/\halpha\ for the two sets of spectra, with each pair of
observations connected.  Most of the 
differences follow the trajectories expected from changes in the nebular
ionization parameter.  We have chosen to analyze the fixed position data, 
in order to take advantage of their higher signal/noise,
and to maintain consistency with the CTIO data, which were 
taken as fixed pointings.  The only exception is the very large 
\hii\ region M8 (S25), which shows a large inconsistency between fixed
and scanned spectra, presumably because the former is 
is strongly influenced by local variations in ionization 
structure.  In that case we use the drift scanned spectrum, which 
samples a region of 10\arcmin\ $\times$ 3\arcmin\ (EW $\times$ NS),
centered on the position listed in Table 1 (see Fig.~2d). 

\footnotetext{The Digitized Sky Surveys were produced at the 
Space Telescope Science Institute under U.S. Government grant NAG W-2166.
The images of these surveys were based on photographic data obtained
using the Oschin Schmidt Telescope at Palomar Mountain and the
UK Schmidt Telescope.  The plates were processed into the present
digital form with the permission of these institutions.}

Exposure times of 300$-$1200 sec were chosen to provide accurate 
spectrophotometry of the bright diagnostic lines (S/N $>$ 20 in most cases).  
The fainter auroral lines such as 
[\ion{O}{3}]\thinspace $\lambda$4363 were not a target of this study, and
in many cases they were too weak in our spectra to be useful.
The spectrophotometric flux calibration was determined using 
standard stars from Massey et al. (1988) for the blue
spectra, and stars from Oke \& Gunn (1983) for the red spectra.  
The observations were made with 2\farcs5 and 4\farcs5 wide slits, oriented
to limit errors from atmospheric dispersion effects to $<$0.05 mag across
the wavelength range.  

\subsection{CTIO Data}

The southern Galactic \hii\/ regions and all of the LMC and SMC 
\hii\ regions were observed in 1987 Jan, 1987 Dec, and 1988 Jan
using the CTIO 1.0 m and 1.5 m telescopes.
At the time that these observations were made a blue-sensitive CCD
detector was not available, so the blue spectra were obtained 
with a 2D FRUTTI photon counting detector system on the 1.0 m Cassegrain
spectrograph.  The instrument was configured
with a 600 g~mm$^{-1}$ grating blazed at 5000 \AA\ and WG-360 blocking
filter, which provided coverage of 3600--7000 \AA\ and a resolution of 
6 \AA\ FWHM when used with an 8\arcsec\ $\times$ 6\arcmin\ slit.
All observations were made with the slit oriented E--W at a fixed position,
usually centered 10\arcsec\ south of the central star.  Due to the limited 
dynamic range of the 2D FRUTTI detector, observations of bright  
regions were made with neutral density filters, and
exposure times for most objects were 900--1800 sec, largely independent 
of surface brightness.  Standard stars from Stone \& 
Baldwin (1983) were observed with a 12\arcsec\ wide slit for flux calibration.

Red spectra were obtained with the Cassegrain spectrograph on 
the CTIO 1.5 m telescope, equipped with an unthinned 385$\times$576
GEC CCD detector.  Two grating settings were used for most of the 
objects.  A 158 g~mm$^{-1}$ grating blazed at 8000 \AA\ in first 
order and a OG-550 blocking filter provided complete coverage of the
5730--10200 \AA\ region, with a resolution of approximately 20 \AA\ FWHM
when used with a 6\arcsec\ $\times$ 6\arcmin\ slit.  
A second set of spectra were obtained
using a 600 g~mm$^{-1}$ grating blazed at 6750 \AA, to cover the
region 5650--6850 \AA\ at 5 \AA\ resolution, again with a 6\arcsec\ slit.
The low-resolution spectra were used primarily to measure the diagnostic
lines longward of 7000 \AA\ (mainly \siii\ and \ariii), while the 
higher-resolution spectra provided the best measurements of the \nii, \sii,
and \ion{He}{1} $\lambda$6678 lines.  Exposure times ranged from 120 to 1800 
sec.  Standard stars from Stone and Baldwin (1983) and Baldwin \& Stone 
(1984) were observed with a 12\arcsec\ slit.  

In order to investigate the local variation in the nebular diagnostic indices
within individual \hii\ regions, we obtained spectra at additional positions
in the Carina and 30 Doradus \hii\ regions, as shown in Figures 2c and 
2g, respectively.  The observations of 30 Dor include a series of five 
positions extending E--W a total of 24\arcmin\ through the center of the \hii\
region.  We also obtained red spectra only at 22 other positions in 30 Dor and 
12 positions in the Orion nebula, to match previous observations at 
3700--7200 \AA\ by Mathis, Chu, \& Peterson (1985) and Peimbert \& 
Torres-Peimbert (1977), respectively.  Charts showing the positions of these
measurements can be found in the original papers.  The red spectra were
used to extend the wavelength coverage of the published measurements to the 
\siii\ lines.  We used the overlapping coverage of the \halpha, 
[\ion{N}{2}], and [\ion{S}{2}] lines to firmly tie in the 
[\ion{S}{3}] fluxes, and confirm that the identical regions were observed. 

\subsection{Data Reduction and Calibration}\label{reduction}

The spectra were reduced using the two-dimensional spectrum reduction
package in IRAF\footnotemark, following conventional procedures.
We limit the discussion here to reductions and tests that were 
unique to this data set.  Spatial distortions in the long-slit spectra were 
removed using comparison lamp exposures.  The distortions were
considerable in the 2D FRUTTI data, and data within 5--10\%\
of the slit edges were excluded in subsequent reductions.
The 2D FRUTTI spectra were checked for deadtime effects in the bright lines,
using short exposures obtained with darker neutral filters, and by 
comparing the observed \oiii\ doublet ratio (the strongest feature in most
raw spectra) to its theoretical value.  Deadtime effects were present in only
a small fraction of the spectra, and in such cases the 
[\ion{O}{3}]\thinspace $\lambda$5007 line was discarded, 
and the scaled flux of [\ion{O}{3}]\thinspace $\lambda$4959 was used instead.
Similar checks were carried out with multiple neutral density
filters on the standard stars, and measurements with any hint of a
saturation problem were discarded.

\footnotetext{IRAF is 
distributed by the National Optical Astronomical Observatories, which
are operated by AURA Inc., under contract with the National Science
Foundation.}  

Calibration of the red CCD spectra in the \siii\ region also required special
care, due to the presence of unresolved telluric water absorption lines, 
To maximize the reliability of the [\ion{S}{3}] measurements we made frequent
observations of subdwarf standard stars located at the same airmass as the 
\hii\/ regions, and used a high-order spline fit to the stellar spectra
to remove the absorption on
wavelength scales of $>$50 \AA\ (exclusive of the regions around the
Paschen emission lines, which are influenced by absorption lines in the
standard stars).  The reliability of this procedure was
confirmed by the degree of consistency of the observed \siii\ doublet ratios
with its theoretical value (2.44), and by the shapes of the 
calibrated continuum spectra in the \hii\ regions themselves.
As a conservative measure we attach a mean uncertainty of 
$\pm$20\%\ to the summed \siii\ fluxes.

The overall precision of the spectrophotometry was determined from 
comparisons of multiple observations of the same objects, intercomparison
of the blue and red spectra in the overlapping wavelength regions, and from
the measured ratios of the [\ion{O}{3}] and [\ion{S}{3}] doublets.
The typical uncertainties for strong lines range from $\pm$5\%\ or better for 
the Bok CCD spectrophotometry (excluding the [\ion{S}{3}]
lines) to $\pm$10\%\ for the CTIO blue spectra (2D FRUTTI) and 
$\pm$20\%\ for the [\ion{S}{3}] fluxes and the CTIO [\ion{O}{2}] fluxes.
External comparisons with published sources such as Dufour (1975), Shaver et
al. (1983), Mathis et al. (1985), and Hunter (1992) show consistency to the 
degree that is expected given the measuring errors above and differences in
positions and aperture sizes.  The only exception is a tendency for the
\ion{He}{1}\ $\lambda$5876 fluxes measured at CTIO (2D FRUTTI) to be 
significantly lower than those from published sources, and consequently 
we attach more weight to the \ion{He}{1}\ $\lambda$6678 (CCD) data in these 
objects. 

\subsection{Emission-Line Measurements}\label{spectra}

A one-dimensional spectrum of each region was extracted 
by summing over the radial extent of the \hii\ region on the slit.
Whenever possible a background sky measured from the outer parts of the slit
was subtracted, but when the \hii\ region filled the slit the sky
was included and subtracted in the fitting of each line flux
(after confirming the absence of sky line contamination). 
Analysis of the extracted 1D spectra followed standard
procedures.  Line fluxes were measured by a direct integration
of the line profile, with subtraction of a linearly interpolated 
continuum, using either the SPLOT task in IRAF or our own software.
The flux scales of the blue and red spectra were tied together using the  
sum of the \halpha\ and \nii\ lines.  

Reddening corrections were derived from the Balmer decrement
(\halpha, \hbeta, \hgamma), the theoretical Balmer line
ratios as calculated by Hummer \& Storey (1987), and the average
interstellar reddening curve from Cardelli, Clayton, \& Mathis (1989).  
When measured electron 
temperatures and densities were available, either from the literature or from 
our spectra, they were used to determine the theoretical Balmer
decrement; otherwise a temperature of 10000 K and density of 100 cm$^{-1}$
were assumed.

Table 2 lists the reddening-corrected line fluxes for the principal
diagnostic lines, as well as the logarithmic extinction at \hbeta\ (column 2).
All fluxes are normalized to $f(H\beta) = 100$.  For \hii\ regions with
multiple measurements we have listed the fluxes for a single representative
position.  Complete line lists,
including the fainter lines and the multiple position measurements are
available from the authors upon request.

\section{ANALYSIS}

Before we specifically address the nebular diagnostics of abundance
and stellar temperature, it is instructive to examine the behavior
of the \hii\ region spectra using the well-known diagnostic diagrams of
Baldwin, Phillips, \& Terlevich (1981), 
and compare them to the spectra of more distant and luminous extragalactic
\hii\ regions.  

To aid in the physical interpretation of these diagnostic
diagrams, we have computed a set of {\sc cloudy} nebular models
(Ferland et al.~1998), using non-LTE stellar atmosphere models by the Munich
group (Pauldrach \etal 1998) as the source of ionizing photons, as
described in BKG.  These will be used later to 
compare the stellar effective temperatures from nebular model fitting
to the actual effective temperatures of the exciting stars.  However in
this section the main use of the models will be to illustrate how 
changes in metal abundance, stellar temperature, or ionization parameter
affect the various line ratios.

\subsection{Diagnostic Diagrams and Ionization Parameter}

Figure 4 shows two of the most commonly used diagnostic diagrams for
extragalactic \hii\ regions and AGNs, with the excitation ratio 
[\ion{O}{3}]$\lambda5007$/H$\beta$ plotted against 
[\ion{N}{2}]$\lambda$6583/H$\alpha$ and \sii/\halpha.
Here and in the following figures we have subdivided the sample:
solid squares mark the calibrating Galactic \hii\ regions ionized
by one (or a few) OB stars; open squares correspond to the more
luminous Galactic \hii\ regions ionized by associations of OB stars
with a range of spectral types; open circles denote the LMC and SMC
regions, and open stars indicate the four nebular ionized by early-type
W-R stars.  For comparison we have plotted with small triangles the same line 
ratios for a large sample of extragalactic \hii\ regions in Sa--Im galaxies, 
from BKG.

Extragalactic \hii\ regions show a tight spectral sequence in both
of these diagrams (McCall, Rybski, \& Shields 1985, Kennicutt \& Garnett
1996), which often is interpreted as evidence that giant \hii\ regions
are predominantly radiation bounded objects (McCall et al. 1985).
Figure 4 shows that the much smaller Galactic \hii\ regions also
lie on this same sequence.  The correspondence is especially tight
in the case of [\ion{O}{3}]/\hbeta\ vs [\ion{N}{2}]/\halpha, where the
sequences are virtually indistinguishable.  The WR nebulae lie
well above the sequence, which reflects the very high effective
temperatures of these stars (\teff\ $\ge$ 55000 K).  The LMC \hii\ regions 
show a slight tendency toward weaker [\ion{N}{2}] emission, which probably
reflects the anomalous N/O ratio in this galaxy (Garnett 1999).

In extragalactic \hii\ regions the tight spectral sequences (Fig.~4)
are mainly interpreted as being abundance sequences (e.g., Searle 1971, 
Shields \& Searle 1978, Pagel \& Edmunds 1978, McCall et al.~1985).
However in our sample the primary variable
along the sequence is not abundance but stellar temperature.  This
is shown in Figure 5, which again shows the [\ion{O}{3}]/\hbeta\ vs 
[\ion{N}{2}]/\halpha\ relation, but this time with the points coded
by stellar temperature (left) and by oxygen abundance (right).  
Although both parameters clearly influence the position of \hii\ regions
along the sequence, variations in \teff\/ clearly dominate in this sample.
The virtual indistinguishability between this \teff\/ sequence
with the extragalactic ``abundance" sequence illustrates 
the degeneracy between temperature and abundance
(and to some degree ionization parameter) in this type of diagnostic
diagram.  In hindsight this may help to explain why it has proven
so difficult to disentagle variations in ionization temperature and
IMF from abundance variations in extragalactic \hii\ regions.

The correspondence between the Galactic and extragalactic sequences
is much poorer for [\ion{O}{3}]/\hbeta\
vs [\ion{S}{2}]/\halpha, with the small Galactic \hii\ regions showing
systematically weaker [\ion{S}{2}] emission (Fig.~4).  A comparison with the
driftscanned spectra in Fig.~3 shows that this is partly due to the 
spatial undersampling of our slit spectra, but the remainder of the 
difference appears to be a systematic difference in mean ionization parameter
between the two samples.  This is quantified in Figure 6, which 
plots the ionization-sensitive ratio \sii/\siii\ as a function of \teff\ 
of the ionizing stars (same symbols as Fig.~4).  The 
The [\ion{S}{2}]/[\ion{S}{3}] ratio is commonly used as a diagnostic of the 
nebular ionization parameter $U$ (D\'{\i}az et al. 1991), defined in this case
as the ratio of ionizing photon to electron densities.  It can be expressed as:
\begin{equation}
U = {{Q_{H^0}} \over {4 \pi {R_S}^2 n c}}
\end{equation}
where $Q_{H^0}$ is the ionizing luminosity of the stars (photons per sec), 
$R_S$ is the Str\"omgren radius of the \hii\ region, $n$ is the 
electron density, and $c$ is the speed of light.  Superimposed on
Fig.~6 are photoionization models for values of 
$\log U$ between $-$1.5 and $-$4.0 (see BKG for details).  
Most of the regions studied here lie in the range of $-3.5 \le \log U 
\le -1.5$, whereas most bright extragalactic \hii\ regions lie in a 
narrower range with $-3.5 \le \log U \le -2.5$ (BKG).  The difference
primarily reflects the larger aperture sizes in extragalactic studies
(typically hundreds of parsecs in projected diameter).  This
difference is not large enough to affect the comparison of
$U$-insensitive indices such as the abundance parameter \x\ and the
nebular hardness parameter \etap\ (\S 4.3), but it may need to be
taken into account when applying more ionization-sensitive indices
such as [\ion{O}{3}]/\hbeta, [\ion{Ne}{3}]/\hbeta, or 
[\ion{O}{3}]/[\ion{N}{2}] (below).

\subsection{Abundance Indices}

The backbone of the extragalactic abundance scale rests on direct
electron temperature based measurements, but in metal-rich \hii\ regions the 
temperature-sensitive auroral lines are unobservable, and most abundance
information in that regime is based on ``empirical" excitation indices 
such as \x\ (Pagel
et al.~1979) or [\ion{O}{3}]$\lambda$5007/[\ion{N}{2}]$\lambda$6583 
(Alloin et al.~1979).
Our data are not of sufficient quality to improve on 
the existing calibrations of these indices, but 
we can use our spatially resolved observations of  
30 Dor, Carina, and Orion to investigate the robustness of the empirical
abundance indices.  Do spectra of different regions in the same object
yield consistent abundances?  

As a test of the \x\ index, the left panel of Figure 7 shows the relation 
between \oiii/\hbeta\ and \oii/\hbeta\ (with linear scales) for the multiple
positions in 30 Dor (open stars), Carina (triangles), and Orion 
(open circles).  The 30 Dor data include spectra from Mathis et al.~(1985)
and the Orion data are all from Peimbert \& Torres-Peimbert (1977).
All three \hii\ regions show a very large range of excitation across
the regions sampled, with both [\ion{O}{3}]/\hbeta\ and [\ion{O}{2}]/\hbeta\
varying by factors of 2--5.  However the sum of the line fluxes (\x) is
relatively constant, as shown by the diagonal lines in Fig. 7,
The constancy of \x\
is especially impressive in 30 Doradus and Carina, where the dispersion
in \x\ transforms to a full range of $<$0.1 dex in O/H.  The \x\ index
is less well-behaved in Orion, though the range in abundances about
the mean value is no larger than the calibration uncertainty of the method.
Our results confirm a similar test applied to 
the giant \hii\ region NGC~604 in M33 (Diaz et al.~1987),
and suggest that even spatially undersampled observations of local 
\hii\ regions may be useful for calibrating \x.

V\'{\i}lchez \& Esteban (1996) and  D\'{\i}az \& P\'erez-Montero (1999) 
have explored the use of the sum of the [\ion{S}{2}] and [\ion{S}{3}]
forbidden-line strengths, \stt\ $\equiv$ (\sii\ $+$ \siii )/\hbeta\
as an empirical abundance index, in analogy to \x.  The main advantage
of this index over \x\ is its monotonic behavior as a function of oxygen
abundance, and its insensitivity to the global degree of ionization of
the nebula, at least for giant extragalactic \hii\ regions.  The right
panel of Fig.~7 shows the behavior of this index for the multiple 
position observations of 30 Dor, Carina, and Orion.  In contrast to  
\x\ (left panel), the \stt\ values show much more scatter and nearly complete
overlap between the three \hii\ regions.  This probably is due to a 
combination of observational error in the \siii\ measurements and
large local variations in \sii/\hbeta, which forms preferentially at
the ionization interfaces and in shocks.  The \stt\ ratio (and hence
the inferred metal abundance) also varies systematically across the nebulae,  
especially in 30 Dor, where the data span the full range of
radius in the \hii\ region.  This suggests that \stt\ should
only be applied to global spectra of \hii\ regions, which sample a 
representative fraction of the S$^{++}$ and S$^+$ emitting volumes.
Despite the considerable scatter in \stt\ within each \hii\ region,
the corresponding dispersion in inferred metal abundances 
is only slightly larger than for \x, due to the relatively
weak $Z$-dependence of \stt\ in this excitation range, if the calibration of 
D\'{\i}az \& P\'erez-Montero (1999) is used.  

Figure 8 shows a similar test of the [\ion{O}{3}]/[\ion{N}{2}] empirical
abundance parameter.  We would expect this ionization-sensitive index
to show a larger dispersion within individual \hii\ regions, and this
is borne out in Fig. 8.  The index shows a dispersion of $\ge$1 dex in all
three \hii\ regions, and the 
inferred abundance values overlap between the objects.  The horizontal lines
indicate the corresponding oxygen abundances in terms of $12 + \log O/H$,
using the calibration of Dutil \& Roy (1999); the inferred abundances
show dispersions of $\pm0.2 - 0.3$ dex in a given \hii\
region, about 2--3 times larger than for \x.  The relevance of these
results for extragalactic applications is unclear, however.  Comparisons of 
oxygen abundances derived from [\ion{O}{3}]/[\ion{N}{2}] and \x\ have
tended to show excellent agreement across a wide range of excitation
(Ryder 1995, Roy \& Walsh 1997), and indeed the former abundances show
a lower scatter, which suggests that the 
[\ion{O}{3}]/\ion{N}{2}] method can be reliable for luminous \hii\ regions
where the spectrophotometric aperture encloses most of the emitting
volume.  However it is also possible that the apparent consistency of
the [\ion{O}{3}/\ion{N}{2}] abundances is the result of a relatively
narrow range of ionization parameters and other nebular properties within
the samples that have been studied to date, and that this consistency
could break down when applied to samples 
of \hii\ regions with systematically different aperture sampling, ionization
structure, or other nebular properties.   

\subsection{Diagnostics of Stellar Temperature and IMF}

Emission-line spectra of \hii\ regions and starbursts are being applied
increasingly to constrain the ages and IMFs in the exciting star clusters
(e.g., Stasi\'nska \& Leitherer 1996, Garcia-Vargas et al. 1997, BKG,
Garnett 2000, and references therein).  These applications rely on the ability 
to use nebular ionization models to characterize the shape and
hardness of the ionizing continuum shortward of the Lyman break.  
In this section we use the sample of calibrating \hii\ regions to
directly assess several hardness parameters, ranging from robust
indices such as the ionized He fraction to crude but commonly used
indices such as the [\ion{O}{3}]/\hbeta\ ratio.  

\subsubsection{He Recombination Lines}

The fractional ionization of He provides a robust measure of stellar
temperature, following methods established originally by Zanstra (1927).
Most of the helium in 
normal \hii\ regions is either neutral or singly ionized,
so the relevant diagnostics are the ratios of the \ion{He}{1}\ and
\ion{H}{1}\ recombination lines.  Figure 9 shows the line ratios
\ion{He}{1}~$\lambda$5876/\hbeta\ and \ion{He}{1}~$\lambda$6678/\halpha\ 
plotted as functions of stellar temperature.  Superimposed are the 
model relations, based on the stellar atmosphere models of BKG and
assuming Case B recombination following Osterbrock (1989).  Models
are plotted for abundances of 1.0 \zsun\ (solid lines) and 0.2 \zsun\
(dashed lines), and for ionization parameter $\log U$ of $-$1.5 
(upper) and $-$4.0 (lower).  Given the relatively narrow range of $U$
in our sample the ionization parameter dependence is unimportant, but
the range of abundance in the sample has some effect on the observed
line ratios (both via line blanketing in the ultraviolet and the He
abundance).  The symbols 
differentiate between calibrating single-star regions, and larger
Galactic and Magellanic Cloud \hii\ regions, following Figure 4.  The
model lines represent the mean He ionization over the entire nebula, so for
the \hii\ regions with partial He ionization we only considered
objects for which we had full or representative spatial coverage of
the nebula.

Figure 9 shows an excellent agreement between the observed levels
of He ionization and the model predictions, especially for 
\ion{He}{1}~$\lambda$6678/\halpha\ where our data are the best
and reddening errors are minimal.  The most discrepant regions are
S275, where our spatial coverage is very incomplete, and S162 (NGC~7635),
which is ionized by an O6.5\thinspace IIIf star.  The latter is the only \hii\ 
region in our sample which is predominantly ionized by a giant star,
but it is unclear whether the discrepancy for this object is coincidental 
or indicative of a general problem with the low-gravity stellar models.
Otherwise the data show that the \ion{He}{1}\ lines provide an excellent
measure of stellar temperature.  Unfortunately this method is only
sensitive when nebular He is partially ionized (35000 K $\le$ \teff\ 
$\le$ 39000 K).  Most extragalactic \hii\ regions 
show nearly complete ionization of helium (\teff\ $\ge$ 40000~K), but the
\ion{He}{1}\ lines do provide important constraints on the ionizing spectrum
in the innermost metal-rich disks of spirals, where many other hardness
indices break down (BKG).  Moreover, Fig.~9 provides independent 
confirmation of the reliability of the Vacca et al. (1996) effective 
temperature scale, at least for 
luminosity class V stars in the \teff\ = 34000 $-$ 39000 K range.  

\subsubsection{Forbidden-Line Excitation Indices}

The simplest and crudest indicators of stellar temperature are based
on single forbidden-line ionization indices, such as [\ion{O}{3}]/\hbeta\
(Chopinet \& Lortet-Zuckermann 1976, Kaler 1978, Copetti, Pastoriza, \&
Dottori 1985, 1986), or analogous infrared
indices such as [\ion{O}{3}]\thinspace 88 $\mu$m/H\thinspace 53$\alpha$ 
or [\ion{N}{3}]\thinspace 57 $\mu$m/H\thinspace 53$\alpha$ (Puxley et al.
1989).  Figure 10 shows the dependence of \oiii/\hbeta\ on stellar
temperature, again with the \hii\ regions subdivided as in Figure 4.
This measure of the excitation shows a tight correlation with
\teff\ (for this sample of \hii\ regions), confirming the good
correlations seen previously by Chopinet \& Lortet-Zuckermann (1976), 
Kaler (1978), and Hunter (1992).  A roughly linear, increasing trend
trend in [\ion{O}{3}]/\hbeta\ with \teff\ is expected, because the
nebular volume containing double-ionized oxygen increases with the
higher ionizing flux emitted by earlier spectral type stars
(Stasi\'{n}ska 1978).  The observed correlation has a remarkably linear
form, extending to the WR stars with temperatures of nearly 70000 K.
The adopted \teff\ scale for the WR stars (Esteban et al.~1993) is 
partly based on nebular modeling, however, so  
strictly speaking the empirical relation in Fig. 10 only applies for
\teff\ $\le$ 50000~K.

Despite the tightness of the correlation in Figure 10, the [\ion{O}{3}]/\hbeta\
ratio must be applied with considerable caution to extragalactic 
\hii\ regions and starbursts, because the ratio is sensitive to 
other variables such as ionization parameter and metal abundance.  
This is illustrated by the four model sequences in Fig. 10, plotted for 
oxygen abundances of 0.2 $Z_\odot$ and 1.0 $Z_\odot$ and 
$\log U$ = $-$2.0 and $-$4.0 (with \teff\/ varying along each sequence).  
The best fitting
models correspond to typical ionization parameters of $\log U \sim -2.5$
and $Z/Z_\odot \sim 0.5 - 1$, consistent with the measured values from 
Fig.~5 and Table 1, respectively.  However the same excitation vs \teff\
relation may not apply in 
external galaxies, where both $Z/Z_\odot$ and $\log$U may be shifted
relative to the typical values in this sample.

We have also investigated the behavior of two other ionization indices
which have been suggested as hardness indicators, \ariii/\halpha\
and \neiii/\oii.  Both are plotted against stellar temperature in
Figure 11.  The figure also shows the temperature dependence of 
\oiii/\hbeta, but plotted this time with a logarithmic scale to
provide a direct comparison with the other two indices.  The behavior of
[\ion{Ar}{3}]/\halpha\ is qualitatively similar to that seen for 
[\ion{O}{3}]/\hbeta, but with a larger scatter and shallower slope 
above \teff $\sim$ 38000 K.  This is not surprising because the lower
ionization potential for Ar$^+$ (27.6 eV vs 35.1 eV for O$^+$) 
is close to the He$^+$ edge at 24.6 eV.  Furthermore the excitation
of [\ion{Ar}{3}] is sensitive to the ionization parameter and abundance
in the same way as [\ion{O}{3}], so it provides a poor substitute 
for the much more robust He$^+$/H$^+$ index discussed earlier.

The bottom panel of Fig. 11 shows the temperature dependence of the
hybrid forbidden-line ratio \neiii/\oii.  Ali et al.~(1991) have shown
that this ratio provides a valuable index for deriving ionization
correction factors in nebular abundance measurements.
Figure 11 shows a strong monotonic dependence of [\ion{Ne}{3}]/[\ion{O}{2}] 
on \teff, which reflects the 41.1 eV ionization potential of Ne$^+$, but
the large scatter in the relation limits its usefulness as a stellar 
thermometer.  The weakness of the [\ion{Ne}{3}] line in metal-rich 
\hii\ regions also limits its value for extragalactic applications. 
A similar plot of [\ion{Ne}{3}]/\hbeta\ shows a nearly
identical scatter, so variations on [\ion{O}{2}] excitation are
not responsible for this dispersion.  Our result is somewhat surprising
in light of the relatively robust behavior of [\ion{Ne}{3}]/\hbeta] in 
the recent analysis of Oey et al. (2000).  Of all of the forbidden-line
ratios that are accessible at visible wavelengths, the familiar
[\ion{O}{3}]/\hbeta\ index appears to be the most useful stellar
temperature indicator, though this sensitivity is accompanied by 
strong dependences on many other nebular properties.

\subsubsection{Composite Ionization Indices: \etap}

The well-known sensitivity of excitation indices such as 
[\ion{O}{3}]/\hbeta\ to nebular ionization structure and abundance
led Mathis (1985) and V\'{\i}lchez \& Pagel (1988) to investigate
the use of composite indices based on the ionization of more than
one element, in an effort to find a more robust UV hardness indicator.
The advent of CCD spectrographs has made the \siii\ doublet accessible,
and with it the possibility of using the $O^{+}/O^{++}$ and 
$S^+/S^{++}$ to contrain \teff\ independently of metal abundance
and ionization parameter.  A convenient hardness index is the 
parameter \etap\ introduced by V\'{\i}lchez \& Pagel (1988):
\begin{equation}
\eta\prime = \frac{[O\thinspace II]\lambda\lambda3726,3729 / \oiii}{\sii / \siii } .
\end{equation}

\noindent
The robustness of \etap\ can be tested directly by plotting the behavior
of [\ion{O}{2}]/[\ion{O}{3}] vs [\ion{S}{2}]/[\ion{S}{3}] for 30 Doradus, 
Orion, and Carina, where
we have spectra at multiple positions.  This is shown in Figure 12a,
along with a series of nebular models from BKG.  
Constant \teff\/ lines are drawn at $Z = 0.5~Z_\odot$  for
\teff\ = 50000, 45000, 40000, and 35000~K (left to right). 
Along each line the ionization
parameter varies from $\log U=-1.5$ (bottom) to $\log U=-4.0$ (top).
The dashed lines show the effects of changing metal
abundance between $Z = 0.2~Z_\odot$ (left) and 1.0 $Z_\odot$ (right,
indicated by small dots), for the same
range of stellar temperatures and a fixed ionization parameter 
$\log U=-2.5$.  For further details on the models see BKG.

For 40000~K $\le$ \teff\ $\le$ 50000~K the lines of constant stellar
temperature have a slope of near unity in this diagram (i.e. constant \etap),
demonstrating the robustness of \etap\ against changes in ionization
parameter, at least in the models.  More impressive yet is the 
distribution of the observed line ratios at the various positions
in 30 Doradus (open stars) and Orion (open circles), which fall on
lines of nearly constant \etap, over ranges of nearly an order of magnitude
in [\ion{O}{2}]/[\ion{O}{3}] or [\ion{S}{2}/[\ion{S}{3}].
The observations of Carina (triangles)
show a large scatter in \etap, and we suspect that this dispersion is real,
reflecting the influence of ionization from stars
with different temperatures across this complex region.  However 
Oey et al. (2000) have also found that \etap\ shows a larger point-to-point
variation in their sample of 4 LMC shell \hii\ regions, and this may
suggest limits to the range of spatial scales and nebular properties
where \etap\ provides a reliable measure of ionizing stellar temperature.

Although the spectra of 30 Dor and Orion are consistent with a nearly
constant value of \etap\ (and hence \teff), Fig.~12a shows a systematic 
offset between the effective temperatures implied by the models
($\sim$39000 K and 37000--38000 K respectively), and those estimated from
the stellar contents of the regions ($\sim$48500 K and 40000 K respectively).
Figure 12b shows the same diagram for the other \hii\ regions in the 
sample (a nebular model for \teff\ = 30000 K is added), and a similar offset
in the stellar temperature scales is apparent, if one compares with the
actual \teff\ values in Table 1.  The \hii\ regions ionized by cooler stars
(\teff\ $<$ 37000 K) seem to show better agreement with the models.
Nevertheless the general consistency between the nebular and spectroscopic 
temperatures is encouraging, confirming and extending earlier results
by Mathis \& Rosa (1991).

Figure~12 impressively demonstrates the robustness of \etap\ against 
systematic variations in $U$, but the models plotted in the figure
suggest a significant dependence of \etap\ on metal abundance.  The
dashed lines show the effects of changing $Z$ from 0.2 \zsun\ (left)
to 1.0 \zsun\ (right).  The $Z$-dependence is relatively weak for
hotter stars (\teff\ $\ge$ 40000 K) and low abundances ($Z \le 0.5$ \zsun), 
but becomes large for cooler stars, where the effects of abundance and stellar
temperature become virtually degenerate, and \etap\ itself
becomes nearly degenerate with [\ion{O}{2}/\ion{O}{3}] and [\ion{O}{3}]/\hbeta.
Inspection of the models suggests that two physical effects contribute
to this abundance sensitivity, increased stellar metal line blanketing in the 
ultraviolet, which softens the ionizing spectrum at a given effective 
temperature, and the increasing dominance of nebular cooling (especially
O$^{++}$) over ionization structure in determining the relative forbidden-line
strengths at high abundance (Shields \& Searle 1978, Oey \& Kennicutt 1993,
Stasinska \& Leitherer 1996).  Thus it is important to confirm
the \etap-derived stellar temperatures for low-excitation nebulae with those
derived from more robust hardness indices, such as the He recombination lines.

The degeneracy between abundance and stellar temperature in model spectra
has hampered previous model-based studies of the metallicity dependence of the 
stellar temperatures and IMFs of extragalactic \hii\ regions (e.g., BKG).
However the data presented in this paper allow to test the correlation between 
\etap\ and \teff\ empirically, independently of the nebular models.
The results are shown in Figure 13; the points show the measured values of
$\log$ \etap\ and \teff, subdivided as in Fig.~4, while the three lines show 
the nebular model relations for 0.2, 0.5, and 1.0 $Z_\odot$ ($\log U = -2.5$
in all three cases).  The \hii\ regions span a range of roughly 2 dex in
$\log$ \etap\ (S237, with no detected [\ion{O}{3}] emission, is not plotted),
and the index shows a monotonic dependence on stellar temperature, albeit
with a considerable scatter.  The general
shape of the $\log \eta\prime$ vs \teff\ relation is consistent with that
predicted by the nebular models; the hatched line shows a quadratic 
fit to the observed \hii\ regions (excluding the WR nebulae, where the
temperature calibration is uncertain), and its general shape is in
good agreement with the model relations for the relevant abundance 
range of $\sim$0.4--1.0 \zsun.  

In contrast to the behavior of our nebular models, our \hii\ region data 
show no evidence for a strong $Z$-dependence of the \etap\ vs \teff\
relation.  In particular, there is no evidence in Fig.~13 for a systematic 
shift between the LMC \hii\ regions
and the more metal-rich (on average) Galactic \hii\ regions, or likewise
for any difference between the most metal-rich and metal-poor Galactic
\hii\ regions.  Within the considerable scatter the trends seen in Fig.~13
are independent of abundance.  The singular exception of the 
SMC \hii\ region N66, which is the most metal-poor region in the sample,
but the \teff\ value of this region almost certainly is underestimated,
due to the presence of the luminous WR star HD 5980 (see the appendix).

We cannot explain why the observed
behavior of \etap\ appears to be more robust than the models would predict,
but we can speculate on the possible reasons.  As pointed out earlier,
our empirical estimates of \teff\ for the \hii\ regions are based on
stellar spectral classifications, using a single, metallicity-independent
spectral type vs \teff\ calibration (Vacca et al. 1996).
We believe that this procedure is justified, because the stellar types
are predominantly assigned on the basis of H, He$^0$, and He$^+$ 
absorption line strengths, which ought be robust against metallicity effects.
However the stellar atmosphere grids which are used in the nebular
models do show considerable changes in ultraviolet line blanketing with
increasing metal abundance, and this may alter the 
ionization temperature at a given fixed effective temperature, in a way
that would not be reflected in the assigned spectral types of 
the stars.  Stated in another way, although we have calibrated \etap\
in terms of stellar effective temperatures in Figure 13, the 
ionization temperatures that are derived from \etap\ may not 
necessarily coincide with \teff\ for abundances that differ substantially
from the local Galactic value.  We return to this point in \S 5.

In summary, our data suggest that \etap\ offers the prospect of estimating 
the characteristic 
stellar temperatures of the ionizing associations in \hii\ regions,
over a relatively large temperature range ($33000 K \le
T_{eff} < 50000 K$), and on a calibration that is directly tied
to the stellar spectral classification system, independent of the
nebular model chain.  However two important limitations of \etap\ must
be recognized before the results in Fig.~13 are applied indiscriminantly to
\hii\ regions in external galaxies.  First, the scatter in the $\log$ \etap\ 
vs \teff\
relation is considerable, amounting to an rms uncertainty of $\pm$3000 K
for any individual region.  Second, photoionization models suggest that 
abundance effects may introduce systematic shifts of a comparable
magnitude in the derived effective temperatures, though the observations
of the calibrating \hii\ regions do not show evidence for shifts of
this magnitude.

\section{Application to Extragalactic \hii\ Regions}

In this section we briefly apply the empirical relations between nebular
hardness properties and stellar temperature from the previous section
to the sample of extragalactic \hii\ regions studied by BKG.
The BKG data are well suited to our purposes because they span a
large range of metal abundance and spiral galaxy type, and they are
based on spectra covering the 3600--9700 \AA\ range, so we can apply
all of the hardness indices discussed earlier, including \etap.  Our
approach is complementary to the model-based studies of V\'{\i}lchez \&
Pagel (1988), Stasi\'nska \& Leitherer (1996), and BKG.

We begin by using the empirical fit to the $\log$ \etap\ vs \teff\ relation
in Fig.~13 (hatched line) to derive stellar temperatures for the BKG
extragalactic \hii\ regions with reliable measurements of \etap.
Fitting a quadratic function to the 
calibrating Galactic \hii\ regions (solid squares) yields:
\begin{equation}
T_{eff} = 50819 - 16485 x + 3778 x^2 
\end{equation}
where $x \equiv \log$ \etap.  Although this relation is based solely on
the smaller calibrating \hii\ regions, where the \teff\ values are the
most reliable, using the entire sample does not change the resulting
relation significantly.  Equation (3) is given primarily so readers
can reproduce and check the results described below; in view of the
caveats discussed in the previous section we {\it strongly} advise
against applying this relation generally to extragalactic regions, 
least of all to luminous starbursts, where the physical conditions may differ
substantially from those of the calibrating \hii\ regions.

With these caveats firmly in mind, the top panel of Figure 14 shows the 
characteristic stellar temperatures \tstar\ and oxygen abundance for 
the BKG sample.  We have deliberately avoided referring to these
as effective temperatures, to emphasize that these are ionization
temperatures which may not coincide precisely with \teff\ outside
the calibrated abundance range.  The abundances 
have been derived using the \x\ empirical method following the calibration of
Edmunds \& Pagel (1984), and should be reliable to within
$\pm$0.2 dex for $Z \le$ \zsun, but less accurate at higher abundances.
However this accuracy is sufficient for testing for general trends in \tstar\ 
with metallicity.  The values of \tstar\ inferred from equation (3)
show a large scatter, consistent with the large scatter observed
in the calibrating sample (Fig.~13), combined with 
any real scatter in \tstar\ at fixed abundance, due to variations
in age and other properties of the ionizing star clusters.  
The large dispersion in Fig. 14a underscores the dangers of applying 
\etap\ or any other forbidden-line hardness parameter to an individual
\hii\ region.  Nevertheless the general behavior of \tstar\ 
should be meaningful.

The top panel of Fig.~14 shows a clear trend in stellar temperature, with the 
typical \teff\/ decreasing from $\sim$55000 K at 0.2 $Z_\odot$ 
($12 + \log O/H = 8.2$)to $\sim$40000 K above solar abundance 
($12 + \log O/H = 8.9$).  This trend is in excellent agreement with the
model-based results of BKG, except for a temperature offset of 
about 5000 K; BKG esitmated a \teff\/ range of 50000 K to 35000 K
over the same abundance range.  In both cases note that most of the \teff\/
change is restricted to abundances below solar.

The measured \ion{He}{1}\ line strengths in the BKG spectra provide another
empirical constraint on the stellar temperatures in metal-rich \hii\ regions.
Unfortunately the He lines are only useful for regions ionized by stars
with \teff\ $\le$ 38000 K, when one allows for observational uncertainties
in the He line fluxes and the dependences on abundance and ionization 
parameter.  Nevertheless we can at least test the consistency of the
\etap\ and He$^+$/He$^0$ temperature scales for the most metal-rich
regions ($12 + \log O/H > 9.0$).  The crosses in Fig.~14 
show the values of \teff\ that are implied by the observed 
\ion{He}{1}\thinspace $\lambda$5876/\hbeta\ ratios and the models
shown in Fig.~9 (the \ion{He}{1}\thinspace $\lambda$6678 line is
usually too faint to provide reliable constraints on \teff), with
with arrows indicating objects for which the He recombination lines only
set lower limits on \teff.  The \ion{He}{1}\ data confirm the general
trend toward lower \tstar\ for $Z >$ \zsun, but with \tstar\ $\sim 
38000 - 39000$ K at the highest abundances, about 2000 K cooler than given
by \etap.  This comparison provides a good illustration of how two different
hardness indices can provide complementary constraints on the stellar
populations in the \hii\ regions.  

As an additional consistency check on these results, the bottom panel of 
Fig.~14 shows the same relation, but in this case with the \tstar\ values 
derived from the average of the \etap\ vs \teff\ calibration (Fig.~13
and eq. [3]), and the relation between [\ion{O}{3}]/\hbeta\ and \teff\
from Fig.~10).  The latter is known to be sensitive
to ionization parameter and to metal abundance itself, but surprisingly
it yields values of \tstar\ that are close to those derived from \etap.
This comparison yields a somewhat lower \tstar\ values at high abundance,
which are in better agreement with the values derived from the \ion{He}{1}\
recombination lines.  However because of the susceptibility of 
[\ion{O}{3}]/\hbeta\ to dependences on parameters other than \teff\
we do not apply this diagnostic further.  We suspect that the general
consistency of the results here are due to the fact that mean abundances
and ionization parameters of the calibrating \hii\ region samples are
typical of the extragalactic \hii\ regions in the BKG sample, but this
does {\it not} imply that the abundance dependence of \tstar\ implied by
the [\ion{O}{3}] excitation is necessarily robust by any means.

In summary the empirically-based stellar temperature indices show
evidence for a systematic decrease in mean stellar temperature
with increasing metal abundance, which appears to qualitatively confirm
previous results based on nebular model fitting (V\'ilchez \& Pagel
1988, BKG).  Several physical effects
can contribute to this softening of the ionizing continuum with 
increasing metallicity: increased line blanketing at constant \teff,
a shift in \teff\ at constant stellar mass, and/or a change in the 
upper mass limit to the IMF.  The models of Stas\'inska \& Leitherer
(1996) suggest that changes in atmospheric blanketing and stellar
evolution can account for most of the changes, and our results do
not contradict this conclusion.  We refer the reader to that
paper and to BKG for a more
complete discussion of the implications of these changes in \teff.

\section{CONCLUSIONS}

In conclusion, we have presented a set of empirical diagnostic line
ratios, calibrated against the effective temperature of the exciting
stars. Our results provide a stronger empirical foundation for many
of the most widely used nebular diagnostics (e.g., \x, \etap), and 
they show much better agreement between observations and models than
often was seen in earlier generations of stellar atmosphere and nebular
photoionization models (cf., Mathis 1985, V\'ilchez \& Pagel 1988).
The empirical relations presented here should be useful in many 
investigations of the
stellar populations of extragalactic \hii\/ regions, where very limited
information is available on the exciting stars, and where an 
estimate of the nebular-based effective temperature is needed, 
even with a restricted wavelength coverage.

We close by reemphasizing the exploratory nature of this study,
and pointing out the most urgently needed improvements before 
one can reliably extract quantitative constraints on the stellar
contents and IMFs of star forming regions from nebular spectra.
On the empirical side, this approach can be much improved if 
accurate ($T_e$-based) metal abundances are obtained for all of the calibrating
\hii\ regions (or their exciting stars), and the effects of effective
temperature, metal abundance, and nebular geometry are accounted for
explicitly in the analysis.  Oey et al. (2000) have demonstrated the
value of obtaining high signal/noise point-by-point measurements of
individual \hii\ regions, and applying this approach to a larger sample
of objects would provide a major improvement over our exploratory 
analysis.  On the theoretical side, more self-consistent
stellar models for the relevant range of abundances, temperatures, and
gravities are the most crucial inputs.  This would also allow for a 
consistent treatment of stellar and nebular abundances in the photoionization
calculations.  We hope that these initial results may stimulate
further progress in these areas.

\acknowledgements

It is a pleasure to thank A.~Pauldrach, R.-P. Kudritzki, and T.~Hoffmann
for sharing their unpublished stellar atmosphere models, and S.~Oey
for a preliminary version of her paper and many discussions about this work. 
This research was supported in part by the National Science Foundation
through grants AST-9421145 and AST-9900789.  

\bigskip

\appendix

\centerline{\bf APPENDIX}

\centerline{\bf EXPLANATORY NOTES ON INDIVIDUAL H\thinspace II REGIONS}

\bigskip

\noindent
S257:  The spectral type adopted is the one predicted by HM90 on the
basis of the observed total nebular ionizing flux.

\noindent
S275 (Rosette): The ionizing cluster is NGC~2244; earliest spectral
type is O4 V((f)); a total of 9 stars earlier than B0.5 has been
included in the computation of \teff.

\noindent
RCW 5 (NGC 2359): A Wolf-Rayet (WN 4) ring nebula. Esteban \etal (1993) 
derive \teff=67$\pm$11 $\times 10^4K$. Given the uncertainties in the
treatment of WR star atmospheres, this object, together with RCW~48,
is included to observe the trends in the nebular diagnostics at very
high \teff\/ values.

\noindent
RCW 16: Walborn (1982) gives two early-type stars in NGC 2467, HD
64568 (O3 V((f*))) and HD 64315 (O6 Vn).

\noindent
RCW 48 (NGC 3199): The second Wolf-Rayet (WN 5) ring nebula in our sample,
with \teff=57$\pm$14 $\times 10^4K$ (Esteban \etal 1993).

\noindent
RCW 53 (Carina):  A total of 36 early-type stars in Tr14 and Tr16 has
been considered, excluding $\eta$ Carinae and one WR star (HD 93162).

\noindent
RCW 57 (NGC~3603): Drissen \etal (1995) identify 11 O-type stars and 3 WN6 stars
which account for at least 80\% of the ionization of the nebula.  The
temperature derived here does not include the WN6 stars, which according
to Esteban et al. (1993) have \teff $\le$ 42000 K.

\noindent
S49 (M16): 20 early-type stars

\noindent
S45 (M17): 12 early-type stars classified from optical and $K$-band
spectroscopy by Hanson, Howarth \& Conti (1997).

\noindent
S100: The spectral type adopted is the one predicted by HM90 on the
basis of the observed total nebular ionizing flux.

\noindent
S158 (NGC 7538):  Ionizing star is CGO 654, with spectral type O7V,
from Crampton, Georgelin, \& Georgelin (1978).

\noindent
S162 (NGC 7635): Ionizing star is BD+60 2522, with spectral type O6.5 IIIf,
from Conti \& Alschuler (1971).  Classified as O6.5 IIIef by Conti \&
Leep (1974).   

\noindent
N11B: We have assumed that the stellar association LH10 is responsible 
for the ionization (26 ionizing stars).

\noindent
N44: The slit was centered on N44B, ionized by LH 47/48.

\noindent
N51D: We have included the 40 early-type stars classified by Oey \&
Smedley (1998) in LH 51 and LH 54.

\noindent
N144: Garmany, Massey \& Parker (1994) list 32 ionizing stars in LH
58. The 3 WR stars were not included.

\noindent
30 Dor: We have included the 64 early-type stars in R136 listed by
Massey \& Hunter (1998), excluding the WR stars.

\noindent
N70: The ionizing cluster is LH 114, studied by Oey (1996).

\noindent
N180: The ionizing cluster has been assumed to be LH 117, studied by
Massey \etal (1989a).

\noindent
N66 (SMC):  The temperature we derived is based on the stellar census
of Massey et al. (1989).  However the calculation does not include
the peculiar WN3+OB binary HD 5980, which according to Massey et al. could 
account for up to half of the ionization of the nebula.  Consequently the
temperature listed in Table 1 is strictly a lower limit.


\newpage
\centerline{\bf FIGURE LEGENDS}

\bigskip

\figcaption{Distribution of oxygen abundances and ionization-weighted stellar
temperatures for the \hii\ region sample.  Open squares denote nebulae
with direct abundance determinations based on electron temperature
measurements.  Filled triangles denote \hii\ regions with empirical
(R$_{23}$) abundance estimates.}

\figcaption{Charts showing the locations of the long-slit spectrophotometric
measurements.  In all cases north is at the top and east is to the left.
The box in M8 (Fig. 2d) indicates the region covered by the drift scan.}

\figcaption{Diagnostic diagrams showing the differences between the fixed
pointing spectra (solid points) and the drift scans (open points) for
the northern sample of Galactic \hii\ regions.}

\figcaption{Comparison of spectral properties of the Galactic calibrating
\hii\ regions (solid squares), large Galactic \hii\ regions (open squares),
LMC and SMC \hii\ regions (open circles), and WR nebulae (open stars).
The small triangles show for comparison the spectral sequence for
extragalactic \hii\ regions in a large sample of galaxies from BKG.}

\figcaption{Relation between [\ion{O}{3}]/\hbeta\ and [\ion{N}{2}]/\halpha\ for
the \hii\ regions in this sample, subdivided by stellar effective
temperature (left) and by nebular oxygen abundance (right).}

\figcaption{The [S II]\lines6716,6731/[S III]\lines9069,9532 vs
\teff\/ diagram for the \hii\ regions.  The symbol coding is the same
as in Fig. 4.  The continuous lines show
nebular models at 0.2 $Z_\odot$, for a variety of ionization
parameter values $\log U$ (as labelled). The dotted lines refer to solar
abundance models at $\log U=-2.0, -2.5$ and $-3.0$.}

\figcaption{(Left): Relation between [\ion{O}{3}]/\hbeta\ and 
[\ion{O}{2}]/\hbeta\
for subregions observed in 30 Doradus (open stars), Carina (solid triangles),
and Orion (open circles).  The dashed lines show constant values of the
empirical abundance parameter \x.  Note the relative constancy of \x\
across the 30 Dor and Carina, despite the large local variations in
excitation.  (Right):  Corresponding relation between [\ion{S}{3}]/\hbeta\ and
[\ion{S}{2}]/\hbeta\ line ratios.  Diagonal lines show constant values
of the abundance index \stt .}

\figcaption{Relation between [\ion{O}{3}]/\hbeta\ and the extragalactic \hii\
region empirical abundance index [\ion{O}{3}]/[\ion{N}{2}], for subregions
observed in 30 Doradus (open stars), Carina (solid triangles), and Orion (open
circles).  The dashed horizontal lines show values of constant oxygen
abundance, using the calibration of Dutil \& Roy (1999).}

\figcaption{Dependence of nebular \ion{He}{1}\ recombination line strengths
(relative to Balmer lines) as a function of stellar effective temperature,
as measured by \ion{He}{1}\thinspace $\lambda$5876 (top) and
\ion{He}{1}\thinspace $\lambda$6678 (bottom).  Both ratios are approximately
proportional to the fraction ionization of He in the nebula.  Symbols
are coded as in Fig. 4.  The lines
show the relations expected from stellar atmosphere and nebular
photoionization models for different abundances, as described in the text.}

\figcaption{Correlation between excitation index [\ion{O}{3}]/\hbeta\ and
stellar effective temperature, with symbols coded as in Fig. 4.  The
four sets of lines show photoionization model sequences for oxygen
abundances of 0.2 and 1.0 $Z_\odot$ and ionization parameters
$\log U = -2.0$ and $-4.0$.}

\figcaption{Dependences of three nebular excitation indices on stellar
effective temperatures.  The symbols are coded as in Figs. 4 and 10.}

\figcaption{Relationship between ionization-sensitive forbidden-line ratios
[\ion{S}{2}]/[\ion{S}{3}] and [\ion{O}{2}]/[\ion{O}{3}].
(a) Spatially resolved observations of the 30 Doradus nebula (open circles),
the Orion nebula (solid squares), and the Carina nebula (open triangles).
The solid lines show photoionization model sequences for stellar
temperatures of 50000 K, 45000 K, 40000 K, and 35000 K (left to right).
(b)  The same line ratios for the full sample of \hii\ regions, with
symbols coded as in Fig. 4.  Solid lines show the same model sequences,
with the addition of a 30000 K model (far right).  Dashed lines show
the effects of increasing the metal abundance, as discussed in the text.}

\figcaption{Dependence of the ionization hardness parameter \etap\
(eq. [2]) and stellar effective for the \hii\ regions in this sample,
with symbols coded as in Fig. 4.  The lines show photoionization model
sequences for oxygen abundances of 0.2, 0.5, and 1.0 $Z_\odot$.
The hatched line shows a quadratic empirical fit to the data.}

\figcaption{The top panel shows the relationship between mean stellar
temperature inferred from the \etap\ parameter (eqs. [2, 3]) and
nebular oxygen abundances, for the BKG sample of extragalactic
\hii\ regions.  The middle panel shows the stellar temperatures
derived from an average of \etap\ and [\ion{O}{3}]/\hbeta, as
described in the text.  Crosses in both panels show values of \tstar\
derived from \ion{He}{1} $\lambda$5876/\hbeta.  Arrows indicate
lower limits to \tstar, corresponding to full He ionization.}



\begin{deluxetable}{lcccccccl}
\scriptsize
\tablecolumns{9}
\tablenum{1}
\tablecaption{H~II region sample\label{sample.table}}
\tablehead{
\colhead{H II Region}     		& 
\colhead{RA (2000)}			&
\colhead{DEC (2000)}   		&
\colhead{T$_*$}				&
\colhead{log Q$_0$}			&
\colhead{12+log(O/H)}			&
\colhead{Ref.}				&
\colhead{Tel.}			&
\colhead{Other ID}		\\
\colhead{(1)}	&
\colhead{(2)}	&
\colhead{(3)}	&
\colhead{(4)}	&
\colhead{(5)}	&
\colhead{(6)}	&
\colhead{(7)}	&
\colhead{(8)}	&
\colhead{(9)}}
\startdata 
\cutinhead{Galaxy}
S212 *    & 04 40 36 & \phantom{-}50 27 46	& 40600	& 49.07 & (8.5) & 1 & Bok &			\nl
S237 *    & 05 31 27 & \phantom{-}34 14 58	& 32900	& 48.21 & (8.9) & 1 & Bok &			\nl
M42 * 	 & 05 35 17 & -05 23 28 		& 40100 & 49.11 & 8.76  & 2,a & CTIO	& Orion, NGC 1976			\nl
S255 *    & 06 12 54 & \phantom{-}17 59 23	& 33300	& 48.02 & 8.32 & 1,b & CTIO & IC 2162		\nl	
S257 *    & 06 12 48 & \phantom{-}18 00 00	& 34100	& 48.37 & 8.26 & 1,b & Bok &			\nl
S271 *    & 06 14 59 & \phantom{-}12 20 16	& 33300	& 48.02 & (8.9) & 1 & Bok & 			\nl
S275     & 06 31 40 & \phantom{-}04 57 48	& 42600	& 49.83 &  8.20 & 3,b & CTIO & Rosette neb.	\nl
S288 *    & 07 08 37 & -04 18 48			& 35900	& 48.46 & (8.7) & 1 & Bok & 			\nl
RCW 6 *   & 07 09 54 & -18 30 21			& 42700	& 49.29 &  8.79 & 4,b & CTIO & S301		\nl
RCW 5    & 07 18 30 & -13 13 48			& 67000	& 48.90 &  8.48 & 5,b & CTIO & NGC 2359, S298	\nl
RCW 8 *   & 07 30 04 & -18 32 13			& 33300	& 48.32 &  8.67 & 1,b & CTIO & S305		\nl
S307 *    & 07 35 33 & -18 45 34			& 33300	& 48.02 & (8.8) & 1 & Bok & 		\nl
RCW 16 *  & 07 52 19 & -26 26 30			& 48700	& 49.95 &  8.56 & 6,b & CTIO & NGC 2467, S311	\nl
RCW 34 *  & 08 56 28 & -43 05 46			& 37200	& 48.64 &  8.90 & 7,b & CTIO & 			\nl
RCW 40 *  & 09 02 21 & -48 41 55			& 38400	& 48.80 &  8.99 & 8,b &	CTIO & 		\nl
RCW 48   & 10 16 33 & -57 56 02			& 57000	& 49.57 & 8.85 & 5,b & CTIO & NGC 3199	\nl
RCW 53   & 10 44 19 & -59 53 21 		& 45700	& 50.91 & 8.49 & 9,b & CTIO & Carina neb., NGC 3372 \nl
NGC 3603 & 11 15 09 & -61 16 17			& 48900	& 50.77 & 8.51 & 10,b & CTIO & RCW 57		\nl
RCW 62   & 11 38 20 & -63 22 22			& 41700	& 50.06 & 8.54 & 11,b & CTIO & $\lambda$ Cen	\nl
M8       & 18 03 37 & -24 23 12			& 41000	& 49.05 & 8.74 & 12,c & Bok & NGC 6523, S25, Lagoon 	\nl
M16      & 18 18 48 & -13 47 00			& 40500	& 50.21 & 8.76 & 13,d & Bok & NGC 6611, S49, Eagle 	\nl
M17      & 18 20 26 & -16 10 36			& 45600	& 50.37 & 8.81 & 14,a & Bok & NGC 6618, S45, Omega	\nl
S99 *     & 20 00 54 & \phantom{-}33 29 48      	& 38400	& 48.80 & (8.7) & 1 & Bok & \nl
S100 *    & 20 01 44 & \phantom{-}33 31 14 	& 46100	& 49.82 & (8.4) & 1 & Bok & 			\nl
S148 *    & 22 56 09 & \phantom{-}58 30 00 	& 33300	& 48.02 & (8.9) & 1 & Bok &			\nl
S152 *    & 22 58 40 & \phantom{-}58 47 01	& 35900	& 48.46 & (8.7) & 1 & Bok & 			\nl
S156 *    & 23 05 59 & \phantom{-}60 15 01	& 40300	& 49.28 & (8.7) & 1 & Bok & IC 1470		\nl
S158 *	 & 23 13 46 & \phantom{-}61 28 21       & 41000 & 49.05 & 8.34 & 8,e & Bok & NGC 7538          \nl
S162 *    & 23 20 44 & \phantom{-}61 11 41       & 41250 & 49.47 & 8.66 & 2,e & Bok & NGC 7635, Bubble          \nl
\cutinhead{LMC}
N11B     & 04 54 49 & -66 25 36			& 45100	& 50.81 & 8.49 & 15,b & CTIO & DEM 34 		\nl
N44      & 05 22 06 & -67 55 00			& 40800	& 50.53 & 8.56 & 16,b & CTIO & DEM 152 		\nl
N138D,B  & 05 24 23 & -68 31 48 		& 67000 & \nodata & 8.50 & b & CTIO & DEM 174			\nl
N51D     & 05 26 14 & -67 30 18			& 39700	& 50.46 & 8.48 & 17,b & CTIO & DEM 192		\nl
N144     & 05 26 33 & -68 51 48			& 41100	& 50.43 & 8.51 & 18,b & CTIO & DEM 199		\nl
N57C     & 05 33 10 & -67 42 48 		& 67000 & \nodata & 8.56 & b & CTIO & NGC 2020, DEM 231	\nl
30 Dor   & 05 38 42 & -69 06 03			& 48500	& 51.63 & 8.41 & 19,d &	CTIO &		\nl
N70      & 05 43 21 & -67 50 48			& 45300	& 50.31 & 8.60 & 20,b & CTIO & DEM 301		\nl
N180     & 05 48 14 & -70 02 00			& 45600	& 50.48 & 8.41 & 21,b & CTIO & DEM 322		\nl
\cutinhead{SMC}
N66      & 00 59 18 & -72 10 48 		& 41900	& 50.72 & 8.22 & 22,d & CTIO & NGC 346		\nl
\enddata
\scriptsize
\tablerefs{Sources of spectral types: 1. Hunter \& Massey 1990; 2. Conti \& Alschuler 1971;
3. Massey etal 1995; 4. Lahulla 1987; 
5. Esteban etal 1993; 6. Walborn 1982; 7. Heydari-Malayeri 1988; 
8. Georgelin etal 1973; 9. Massey \& Johnson 1993; 10. Drissen etal 1995;
11. Walborn 1987; 12. Lada etal 1976; 13. Hillenbrand etal 1993; 
14. Hanson etal 1997; 15. Parker etal 1992; 16. Oey \& Massey 1995;
17. Oey \& Smedley 1998; 18. Garmany etal 1994; 19. Massey \& Hunter 1998;
20. Oey 1996; 21. Massey etal 1989a; 22. Massey etal 1989b.\newline
Sources of abundances: a. Peimbert etal 1993; b. This paper;
c. Melnick etal 1989; d. Shaver etal 1983; e. Talent \& Dufour 1979.
}
\end{deluxetable}


\begin{deluxetable}{lcccccccccc}
\scriptsize
\tablecolumns{11}
\tablewidth{0pt}
\tablenum{2}
\tablecaption{Reddening-corrected line fluxes}

\tablehead{
\colhead{HII Region}     	& 
\colhead{c}			&
\colhead{[O II]}   		&
\colhead{[Ne III]}		&
\colhead{[O III]}	   	&
\colhead{He I}   		&
\colhead{[N II]}   		&
\colhead{He I}			&
\colhead{[S II]}   		&
\colhead{[Ar III]}		&
\colhead{[S III]}   		\\
\colhead{} 			&
\colhead{}			&
\colhead{3727} 			&
\colhead{3869}			&
\colhead{4959+5007} 		&
\colhead{5876} 			&
\colhead{6548+6583} 		&
\colhead{6678}			&
\colhead{6716+6731} 		&
\colhead{7135}			&
\colhead{9069+9532}		\\
\colhead{(1)}	&
\colhead{(2)}	&
\colhead{(3)}	&
\colhead{(4)}	&
\colhead{(5)}	&
\colhead{(6)}	&
\colhead{(7)}	&
\colhead{(8)}	&
\colhead{(9)}	&
\colhead{(10)}	&
\colhead{(11)}}
\startdata 
S212     & 1.25 & 332 &    0 &  249 &    13 &    68 &    4 &   30 &   123 &    87  \nl
S237     & 0.99 & 305 &    3 &    0 &     0 &   160 &    0 &   69 &    0 &    54  \nl
M42      & 0.44 & 182 &    8 &  263 &    11 &    92 &    3 &   14 &   10 &   118  \nl
S255     & 1.53 & 407 &   10 &   20 &     5 &   152 &    1 &   64 &    4 &    73  \nl
S257     & 0.88 & 208 &    0 &   22 &     5 &   125 &    2 &   58 &    4 &    79  \nl
S271     & 1.29 & 319 &    0 &    8 &     2 &   140 &    2 &   48 &    3 &    29  \nl
S275     & 1.03 & 182 &    5 &  203 &    11 &    41 &    2 &   22 &    8 &    68  \nl
S288     & 1.09 & 291 &   14 &  171 &    10 &    77 &    2 &   20 &    7 &    94  \nl
RCW 6    & 0.79 & 278 &    5 &  186 &    13 &    94 &    4 &   50 &   11 &   189  \nl
RCW 5    & 1.14 & 346 &  137 & 1431 &    16 &    85 &    7 &   75 &   27 &   259  \nl
RCW 8    & 1.83 & 423 &    0 &    7 &     3 &   144 &    1 &   53 &    1 &    86  \nl
S307     & 1.42 & 300 &    0 &   43 &     6 &   129 &    1 &   39 &    4 &    90  \nl
RCW 16   & 0.67 & 180 &   16 &  384 &    12 &    40 &    4 &   15 &   13 &   194  \nl
RCW 34   & 1.91 & 250 &   10 &  122 &    11 &   103 &    4 &   31 &    0 &   185  \nl
RCW 40   & 1.30 & 380 &    0 &   64 &    10 &   152 &    5 &   43 &   11 &   163  \nl
RCW 48   & 1.67 & 241 &   56 &  783 &    10 &    83 &    4 &   47 &   20 &   188  \nl
RCW 53   & 0.87 & 405 &   10 &  210 &    10 &   102 &    4 &   32 &   12 &   126  \nl
NGC 3603 & 2.80 & 172 &   47 &  584 &    12 &    27 &    5 &    6 &   19 &   163  \nl
RCW 62   & 1.02 & 293 &    3 &  147 &    10 &    60 &    4 &   13 &   10 &   118  \nl
M8       & 0.64 & 249 &    4 &  147 &    11 &   132 &    3 &   18 &   12 &   135  \nl
M16      & 0.89 & 212 &    3 &  118 &    12 &   128 &    3 &   30 &   10 &    89  \nl
M17      & 1.24 & 132 &   20 &  437 &    14 &    53 &    4 &   13 &   16 &   121  \nl
S99      & 1.84 & 194 &    6 &  230 &    14 &    56 &    4 &   22 &   11 &   118  \nl
S100     & 1.99 & 101 &   32 &  560 &    16 &    31 &    5 &    9 &   16 &   138  \nl
S148     & 1.28 & 317 &    0 &   10 &     5 &   187 &    1 &   60 &   10 &   189  \nl
S152     & 1.59 & 325 &    0 &  113 &    11 &   118 &    3 &   28 &   11 &   146  \nl
S156     & 1.43 & 297 &    2 &  121 &    12 &   121 &    4 &   25 &   13 &   148  \nl
S158	 & 1.70 & 168 &  15  &  403 & \nodata  &    38  &    3  &   16  &   13  &   120     \nl
S162	 & 0.77 & 226 &   2  &  172 & \nodata  &    98    &    2  &   21  &    9.4  &   160     \nl
N11B     & 0.39 & 310 &   19 &  406 &    11 &    25 &    4 &   22 &   11 &   103  \nl
N44      & 0.22 & 413 &   12 &  297 &     8 &    36 &    3 &   44 &    9 &   110  \nl
N138D,B  & 0.25 & 299 &   69 & 1048 &    13 &    19 &    0 &   52 &   19 &   169  \nl
N51D     & 0.37 & 470 &    8 &  242 &    12 &    26 &    3 &   40 &    7 &    75  \nl
N144     & 0.30 & 293 &   26 &  419 &     9 &    26 &    4 &   33 &   11. &   113  \nl
N57C	 & 0.28 & 226 &   58 &  891 &    14 &    27 &    4 &   47 &   14 &   109  \nl
30 Dor   & 0.62 & 245 &   40 &  569 &    11 &    15 &    4 &   19 &   11 &   110  \nl
N70      & 0.25 & 478 &   13 &  243 &    12 &    36 &    3 &   75 &   11 &    82  \nl
N180     & 0.38 & 328 &   20 &  373 &    11 &    25 &    4 &   28 &    9 &    86  \nl
N66      & 0.28 & 122 &   49 &  752 &     9 &     6 &    3 &   14 &    9 &    58  \nl
\enddata
\end{deluxetable}


\begin{references}

\reference{} Ali, B., Blum, R.D., Bumgardner, T.E., Cranmer, S.R., 
Ferland, G.J, Haefner, R.I., \& Tiede, G.P. 1991, \pasp, 103, 1182

\reference{} Alloin, D., Collin-Soufrin, S., Joly, M., \& Vigroux, L. 1979,
  \aap, 78, 200

\reference{} Baldwin, J.A., Phillips, M.M., \& Terlevich, R. 1981,
\pasp, 93, 5

\reference{} Baldwin, J.A., \& Stone, R.P.S. 1984, \mnras, 206, 241

\reference{} Bresolin, F., Kennicutt, R.C., \& Garnett, D. R. 1999,
\apj, 510, 104

\reference{}  Cardelli, J. A., Clayton, G. C., \& Mathis, J. S. 1989, \apj,
  345, 245

\reference{} Chopinet, M., \& Lortet-Zuckermann, M.C. 1976, \aaps, 25, 179

\reference{} Chopinet, M., Georgelin, Y.M., \& Lortet-Zuckermann,
M.C. 1973, \aap, 29, 225

\reference{} Crampton, D., Georgelin, Y. M., \& Georgelin, Y. P. 1978, 
  \aap, 66, 1

\reference{} Conti, P. S., \& Alschuler, W. R. 1971, \apj, 170, 325

\reference{} Conti, P. S., \& Leep, E. M. 1974, \apj, 193, 113

\reference{} Copetti, M.V., Pastoriza, M.G., \& Dottori, H.A. 1985, \aap,
152, 427

\reference{} Copetti, M.V., Pastoriza, M.G., \& Dottori, H.A. 1986, \aap,
156, 111

\reference{} D\'{\i}az, A. I., Terlevich, E., Pagel, B. E. J., V\'{\i}lchez, J. M., \& Edmunds, M. G. 1987, \mnras, 226, 19

\reference{} D\'{\i}az, A. I., Terlevich, E., V\'{\i}lchez, J. M., Pagel, B. E. J., \& Edmunds, M. G. 1991, \mnras, 253, 245

\reference{} D\'{\i}az, A. I., \& P\'erez-Montero, E. 1999, \mnras, in press

\reference{} Dinerstein, H.L. 1996, in ASP Conf. Ser. 99, Cosmic
Abundances, ed. S.S. Holt \& G. Sonneborn (San Francisco: ASP), 337

\reference{} Drissen, L., Moffat, A.F.J., Walborn, N.R., \& Shara,
M.M. 1995, \aj, 110, 2235

\reference{} Dufour, R.J. 1975, \apj, 195, 315

\reference{} Dutil, Y., \& Roy, J.-R. 1999, \apj, 516, 62

\reference{} Edmunds, M.G., \& Pagel, B.E.J. 1984, \mnras, 211, 507

\reference{} Esteban, C., Smith, L.J., Vilchez, J.M., \& Clegg,
R.E.S. 1993, \aap, 272, 299

\reference{} Evans, I. N., \& Dopita, M. A. 1985, \apjs, 58, 125

\reference{} Ferland, G. J., Korista, K. T., Verner, D. A., Ferguson, J, W., Kingdon, J. B., \& Verner, E. M. 1998, \pasp, 110, 761

\reference{} Garc\'{\i}a-Vargas, M. L., Gonz\'{a}lez-Delgado, R. M., P\'{e}rez, E., Alloin, D., D\'{\i}az, A., \& Terlevich, E. 1997, \apj, 478, 112

\reference{} Garmany, C.D., Massey, P., \& Parker, J.W. 1994, \aj,
108, 1256

\reference{}  Garnett, D.R. 1999, in IAU Symposium 190, New Views of the
Magellanic Clouds, ed. Y.-H. Chu, N. Suntzeff, J. Hesser, \& D. Bohlender,
(San Francisco: ASP), in press

\reference{} Garnett, D.R. 2000, in Spectrophotometric Dating of Stars
and Galaxies, ed. I. Hubeny, S. Heap, \& R. Cornett (San Francisco: ASP),
in press

\reference{} Georgelin, Y.M., Georgelin, Y.P., \& Roux, S. 1973, \aap, 
25, 337

\reference{} Hanson, M.M., Howarth, I.D., \& Conti, P.S. 1997, \apj,
489, 698

\reference{} Heydari-Malayeri, M. 1988, \aap, 202, 240

\reference{} Hillenbrand, L. A., Massey, P., Strom, S.E., \& Merrill,
K.M. 1993, \aj, 106, 1906

\reference{} Hummer, D.G., \& Storey, P.J. 1987, \mnras, 224, 801

\reference{} Hunter, D.A. 1992, \apjs, 79, 469

\reference{} Hunter, D.A., \& Massey, P. 1990, \aj, 99, 846 (HM90)

\reference{} Kaler, J.B. 1978, \apj, 220, 887

\reference{} Kaler, J. B., \& Jacoby, G. H. 1989, \apj, 345, 871

\reference{} Kennicutt, R.C. 1984, \apj, 287, 116


\reference{} Lada, C.J., Gull, T.R., Gottlieb, C.A., \& Gottlieb,
E.W. 1976, \apj, 203, 159

\reference{} Lahulla, J.F. 1987, \aj, 94, 1062

\reference{} Leitherer, C., \& Heckman, T. 1995, \apjs, 96, 9L

\reference{} Leitherer, C. et al. 1999, \apjs, 123, 3

\reference{} Massey, P., Garmany, C.D., Silkey, M., \&
DeGioia-Eastwood, K. 1989a, \aj, 97, 107

\reference{} Massey, P., \& Hunter, D.A. 1998, \apj, 493, 180

\reference{} Massey, P., \& Johnson, J. 1993, \aj, 105, 980

\reference{} Massey, P., Johnson, K.E., \& DeGioia-Eastwood, K. 1995, \apj, 454, 151

\reference{} Massey, P., Parker, J.W., \& Garmany, C.D. 1989b, \aj, 98, 
1305

\reference{} Massey, P., Strobel, K., Barnes, J.V., \& Anderson, E. 1988,
\apj, 328, 315

\reference{} Mathis, J. S. 1985, \apj, 291, 247

\reference{} Mathis, J. S., Chu, Y.-H., \& Peterson, D. E. 1985, \apj, 292, 155

\reference{} Mathis, J. S., \& Rosa, M. R. 1991, \aap, 245, 625

\reference{} McCall, M. L., Rybski, P. M., \& Shields, G. A. 1985, \apjs, 57, 1

\reference{} Melnick, J., Tapia, M., \& Terlevich, R. 1989, \aap, 213, 
89

\reference{} Oey, M.S. 1996, \apj, 465, 231

\reference{} Oey, M.S., \& Kennicutt, R.C. 1993, \apj, 411, 137

\reference{} Oey, M.S., \& Massey, P. 1995, \apj, 452, 210

\reference{} Oey, M.S., \& Smedley, S.A. 1998, \aj, 116, 1263

\reference{} Oey, M.S., Dopita, M.A., Shields, J.C., \& Smith, R.C.
2000, \apj, submitted

\reference{} Oke, J. B., \& Gunn, J. E. 1983, \apj, 266, 713

\reference{} Osterbrock, D. E. 1989, {\it The Astrophysics of Gaseous Nebulae and Active Galactic Nuclei} (Mill Valley, CA: University Science Books)

\reference{} Pagel, B. E. J., Edmunds, M. G., Blackwell, D. E., Chun, M. S., \& Smith, G. 1979, \mnras, 189, 95

\reference{} Pagel, B.E.J., \& Edmunds, M.G. 1978, \mnras, 185, 77


\reference{} Parker, J.W., Garmany, C.D., Massey, P., \& Walborn,
N.R. 1992, \aj, 103, 1205

\reference{} Pauldrach, A. W. A., Lennon, M., Hoffmann, T. L., Sellmaier, F., Kudritzki, R.-P., \& Puls, J. 1998, {\it Second Boulder-Munich Workshop on Hot Stars}, p. 258

\reference{} Peimbert, M., \& Torres-Peimbert, S. 1977, \mnras, 179, 217

\reference{} Peimbert, M., Torres-Peimbert, S., \& Dufour, R.J. 1993,
\apj. 418

\reference{} Puxley, P. J., Brand, P. W. J. L., Moore, T. J. T., 
Mountain, C. M., Nakai, N., \& Yamashita, T. 1989, \apj, 345, 163

\reference{} Rieke, G.H., Loken, K., Rieke, M.J., \& Tamblyn, P. 1993,
\apj, 412, 99

\reference{} Roy, J.-R., \& Roy, J.R. 1997, \mnras, 288, 715

\reference{} Ryder, S. D. 1995, \apj, 444, 610

\reference{} Schaerer, D., \& de Koter, A., 1997, \aap, 322, 598

\reference{} Searle, L. 1971, \apj, 168, 327

\reference{} Shaver, P.A., McGee, R.X., Newton, L.M., Danks, A.C., \&
Pottasch, S.R. 1983, \mnras, 204, 53



\reference{} Shields, G. A., \& Searle, L. 1978, \apj, 222, 821

\reference{} Shields, G. A. 1990, \araa, 28, 525

\reference{} Stasi\'{n}ska, G. 1978, \aaps, 32, 429

\reference{} Stasi\'{n}ska, G., \& Leitherer, C. 1996, \apjs, 107, 661

\reference{} Stone, R.P.S., \& Baldwin, J.A. 1983, \mnras, 204, 347

\reference{} Talent, D. L., \& Dufour, R. J. 1979, \apj, 233, 888

\reference{} Vacca, W. D., Garmany, C. D., \& Shull, J. M. 1996, \apj, 460, 914

\reference{} Vila-Costas, M.B., \& Edmunds, M.G. 1992, \mnras, 259, 121

\reference{} V\'{\i}lchez, J. M., \& Pagel, B. E. J. 1988, \mnras, 231,257


\reference{} V\'{\i}lchez, J. M., \& Esteban, C. 1996, \mnras, 280, 720

\reference{} Walborn, N.R. 1982, \aj, 87, 1300

\reference{} Walborn, N.R. 1987, \aj, 93, 868

\reference{} Zanstra, H. 1927, \apj, 65, 50

\reference{} Zaritsky, D., Kennicutt, R. C., \& Huchra, J. P. 1994, \apj, 420, 87

\end{references}
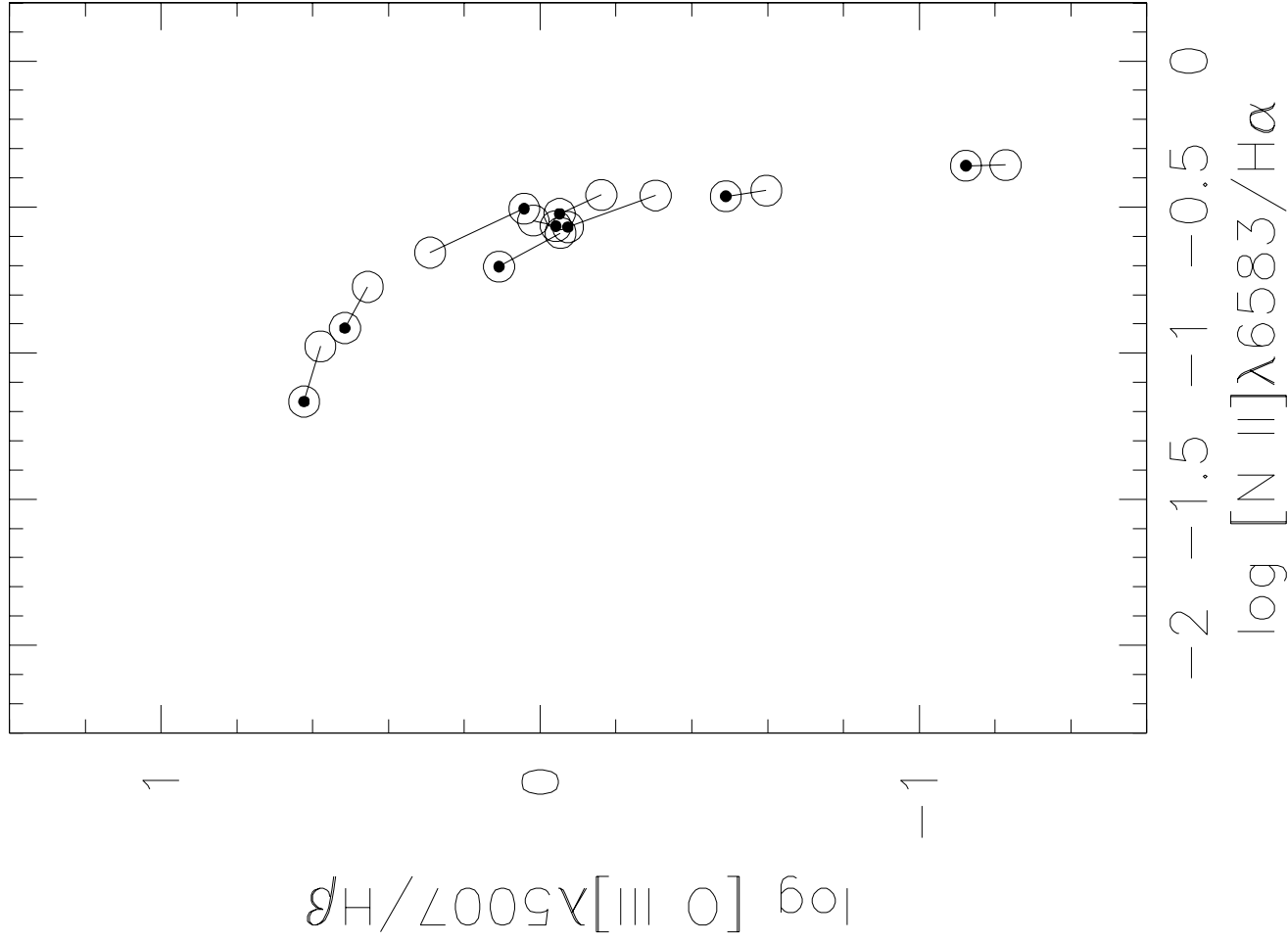
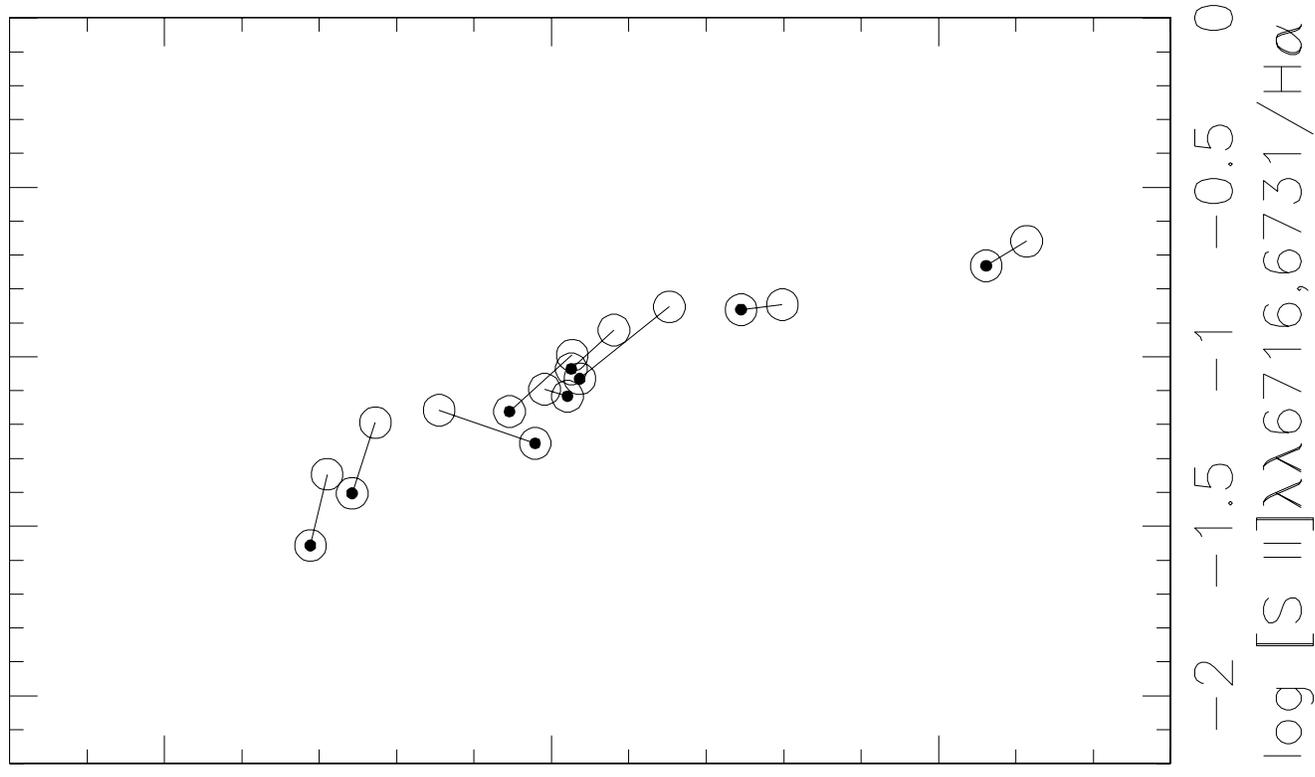
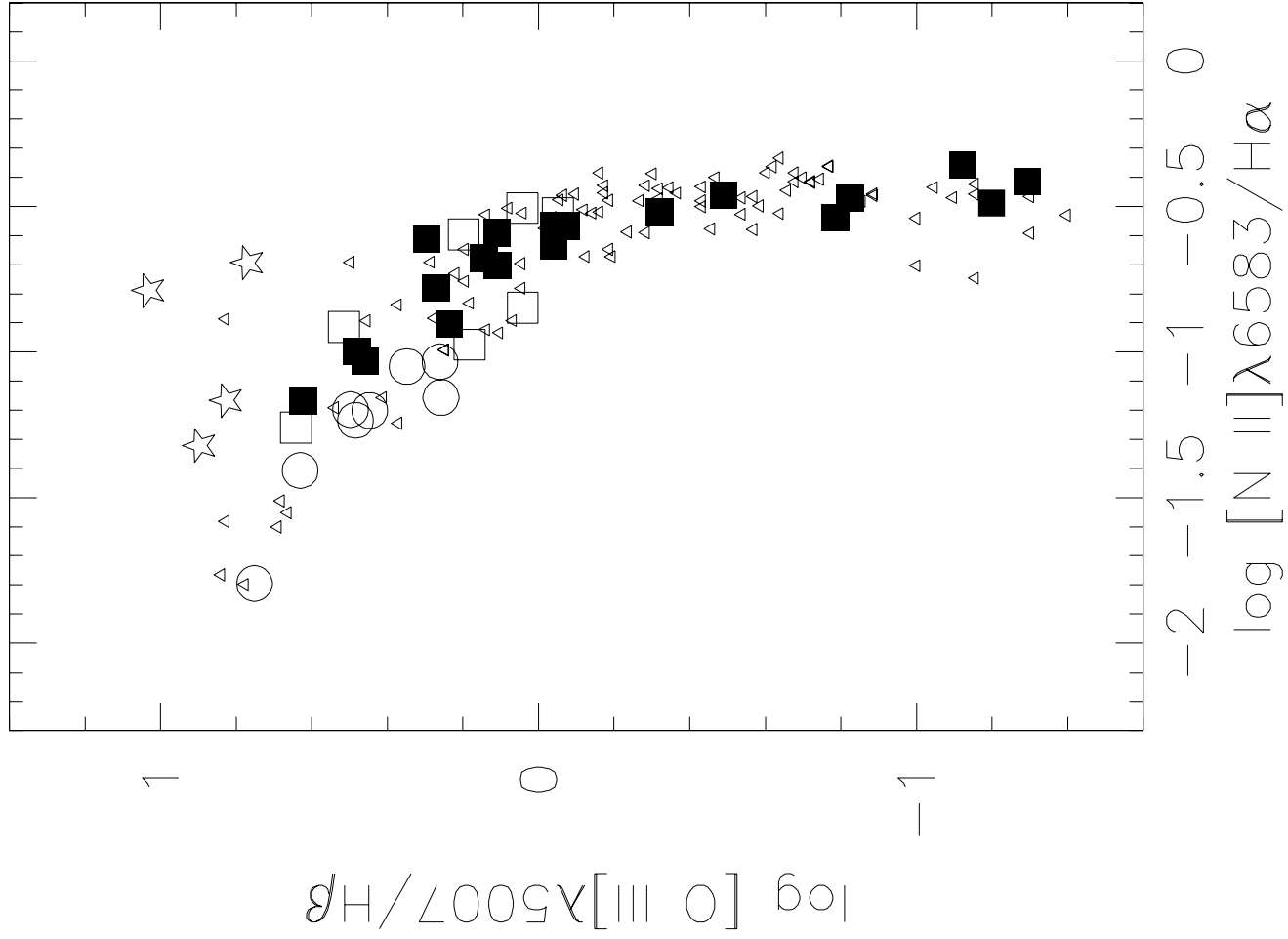
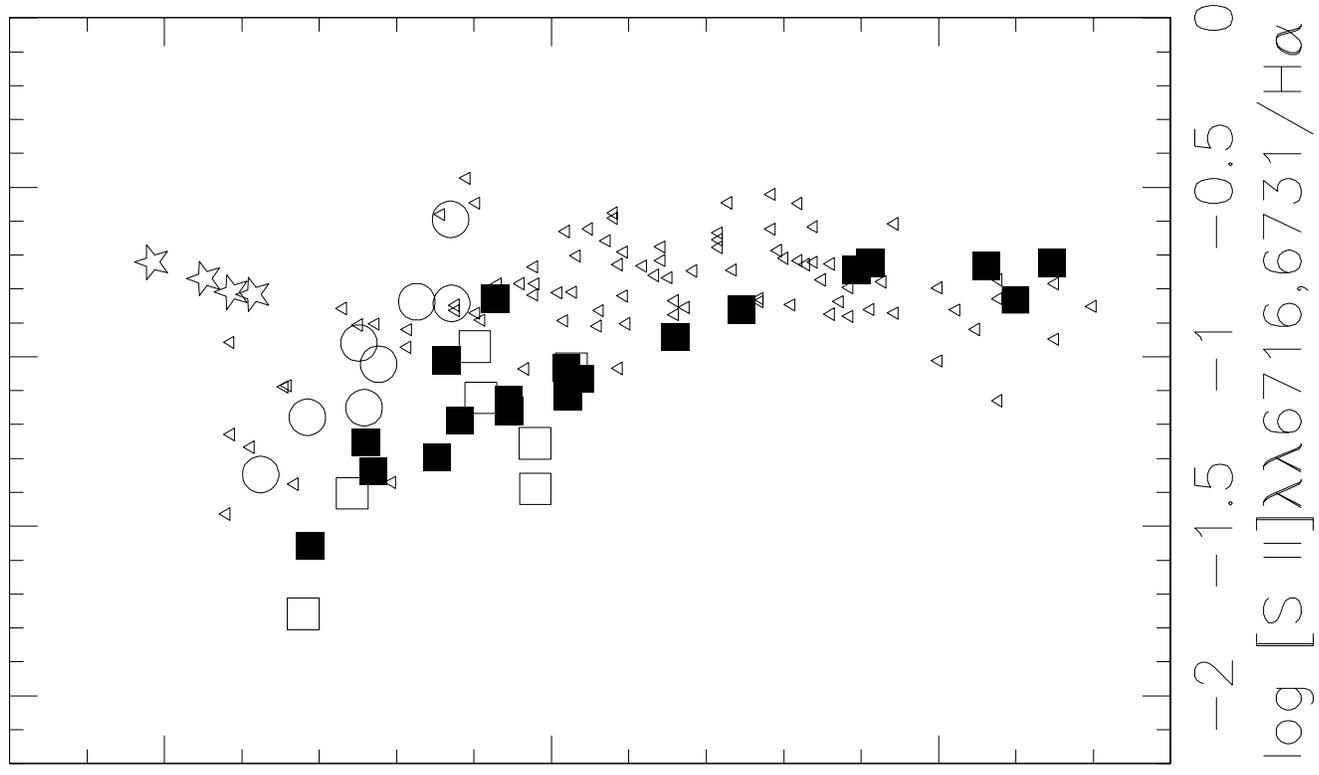
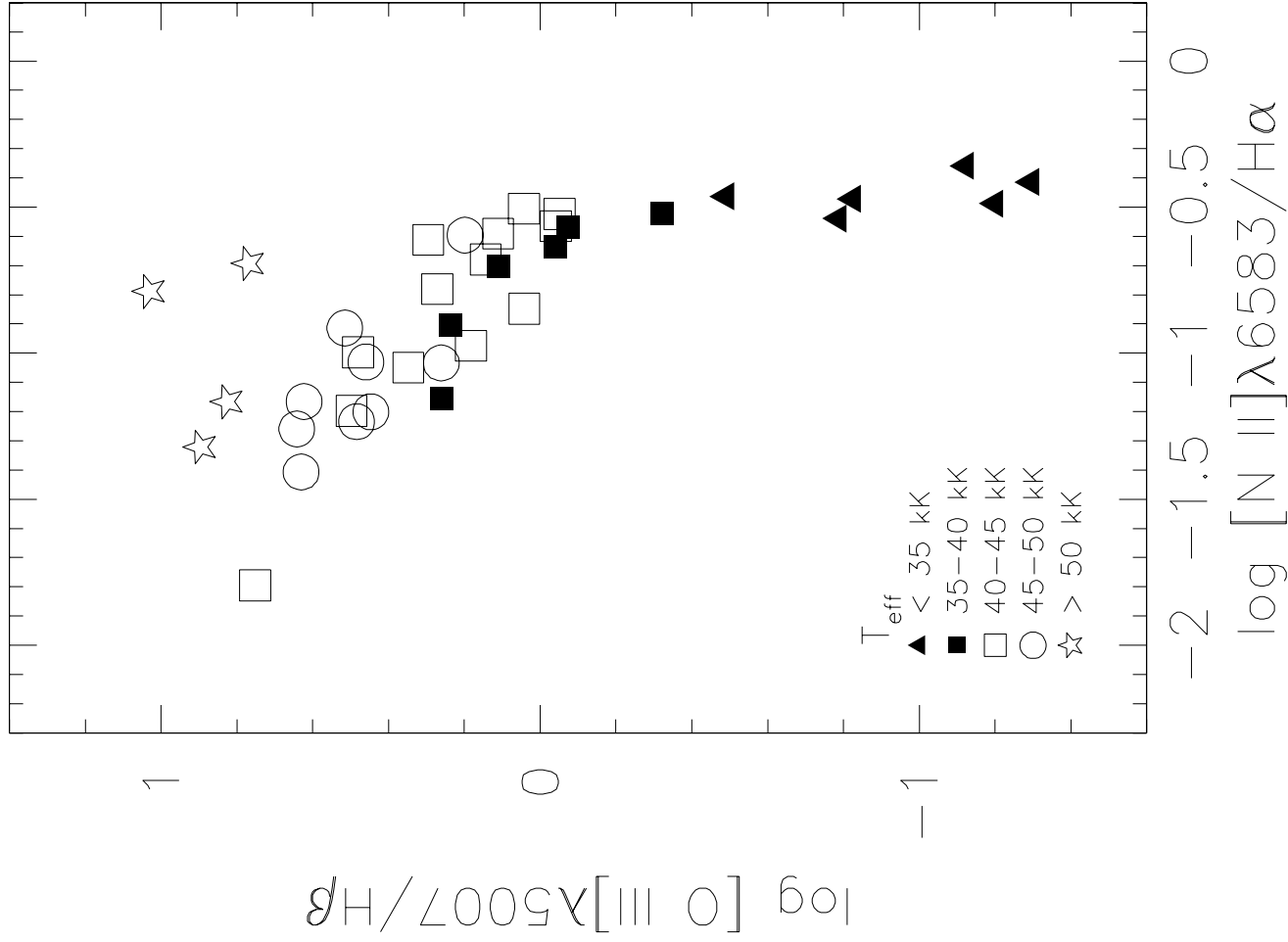
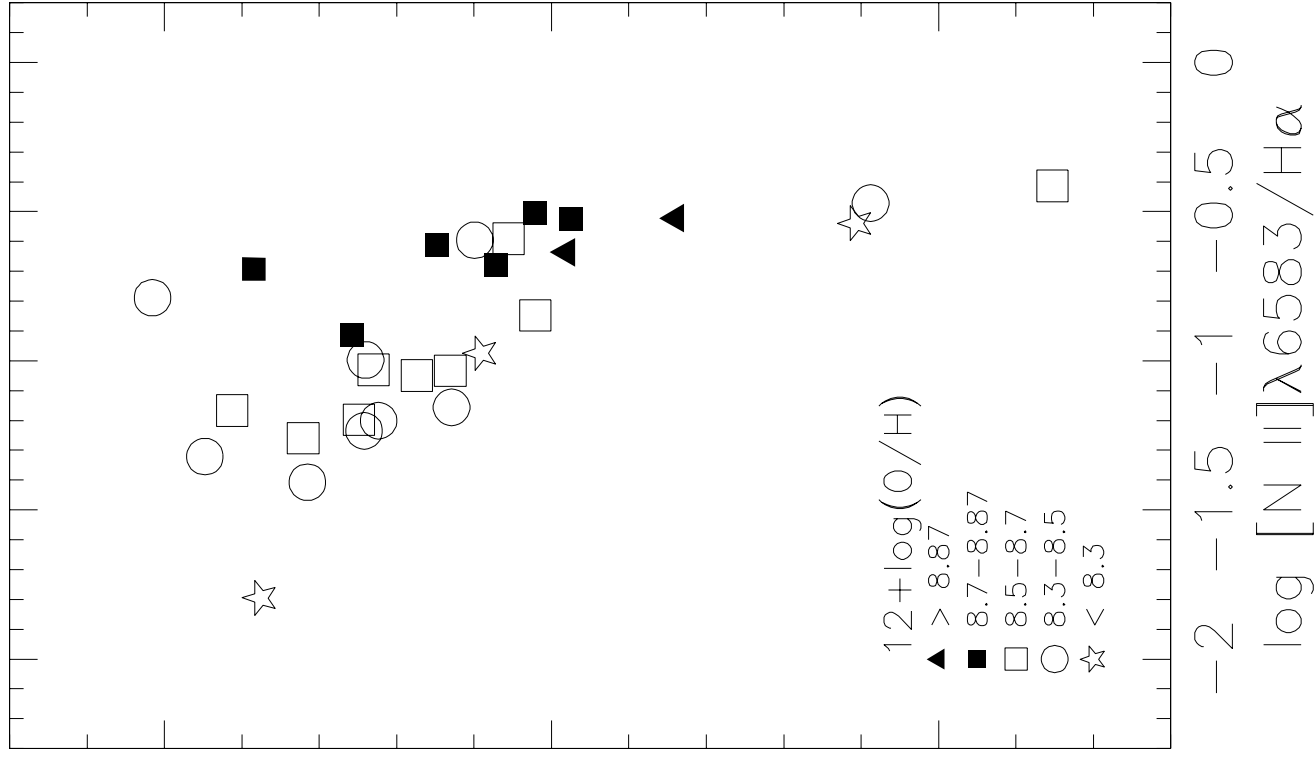
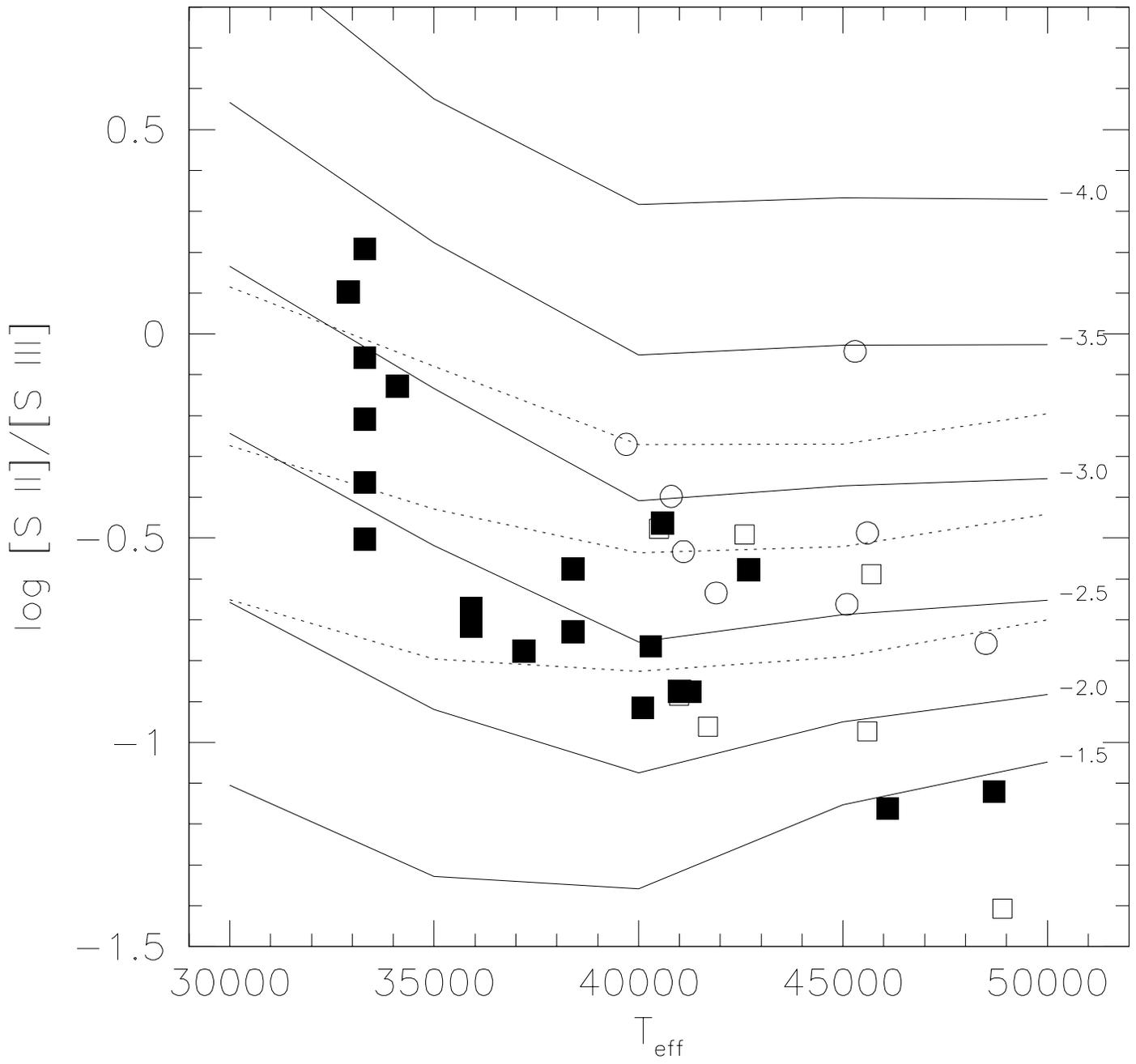
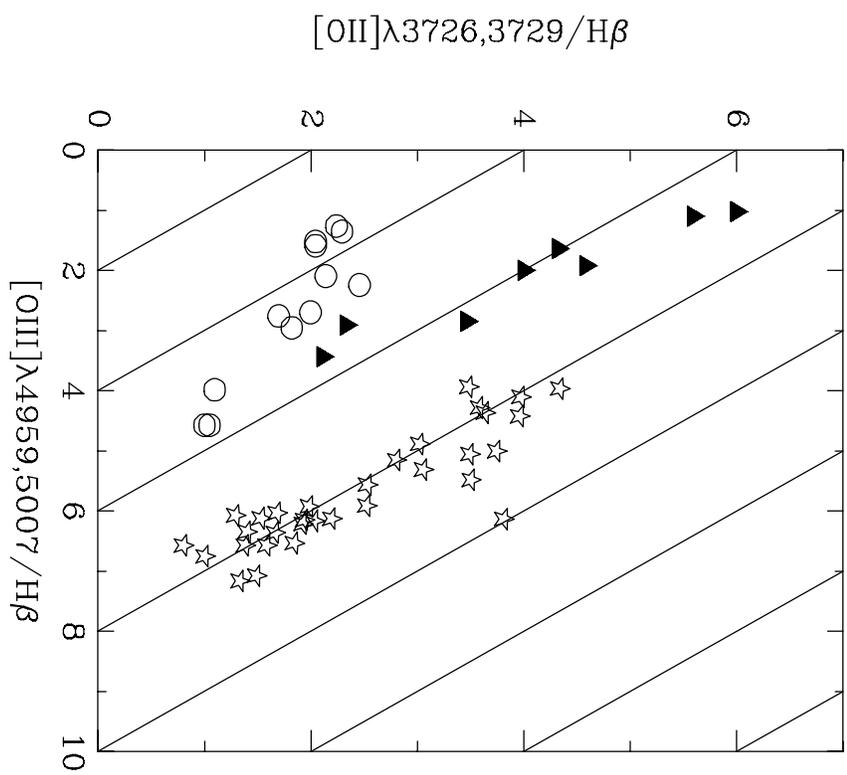
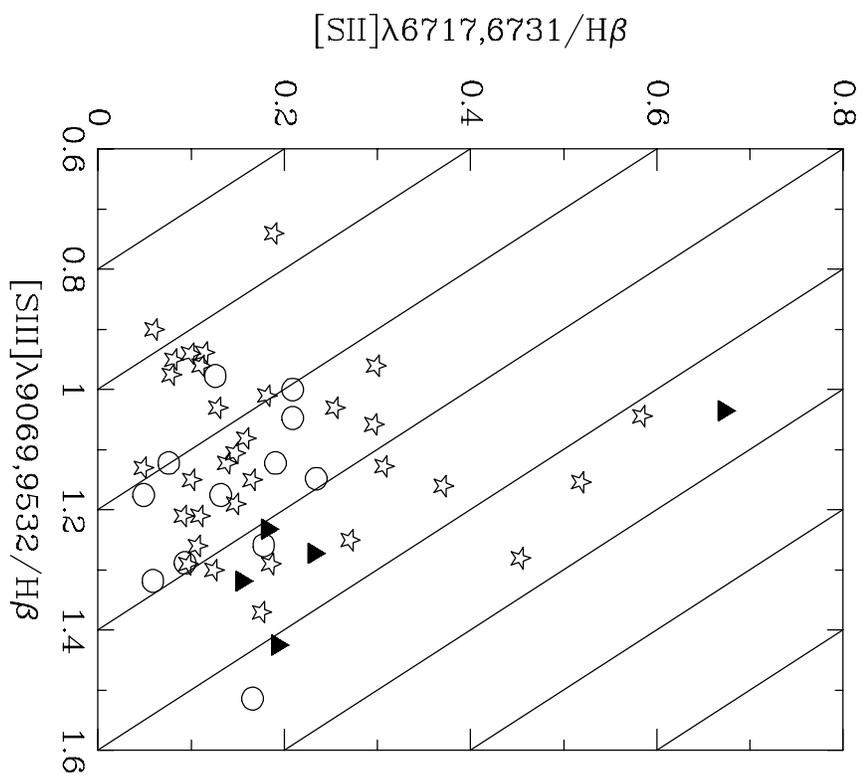
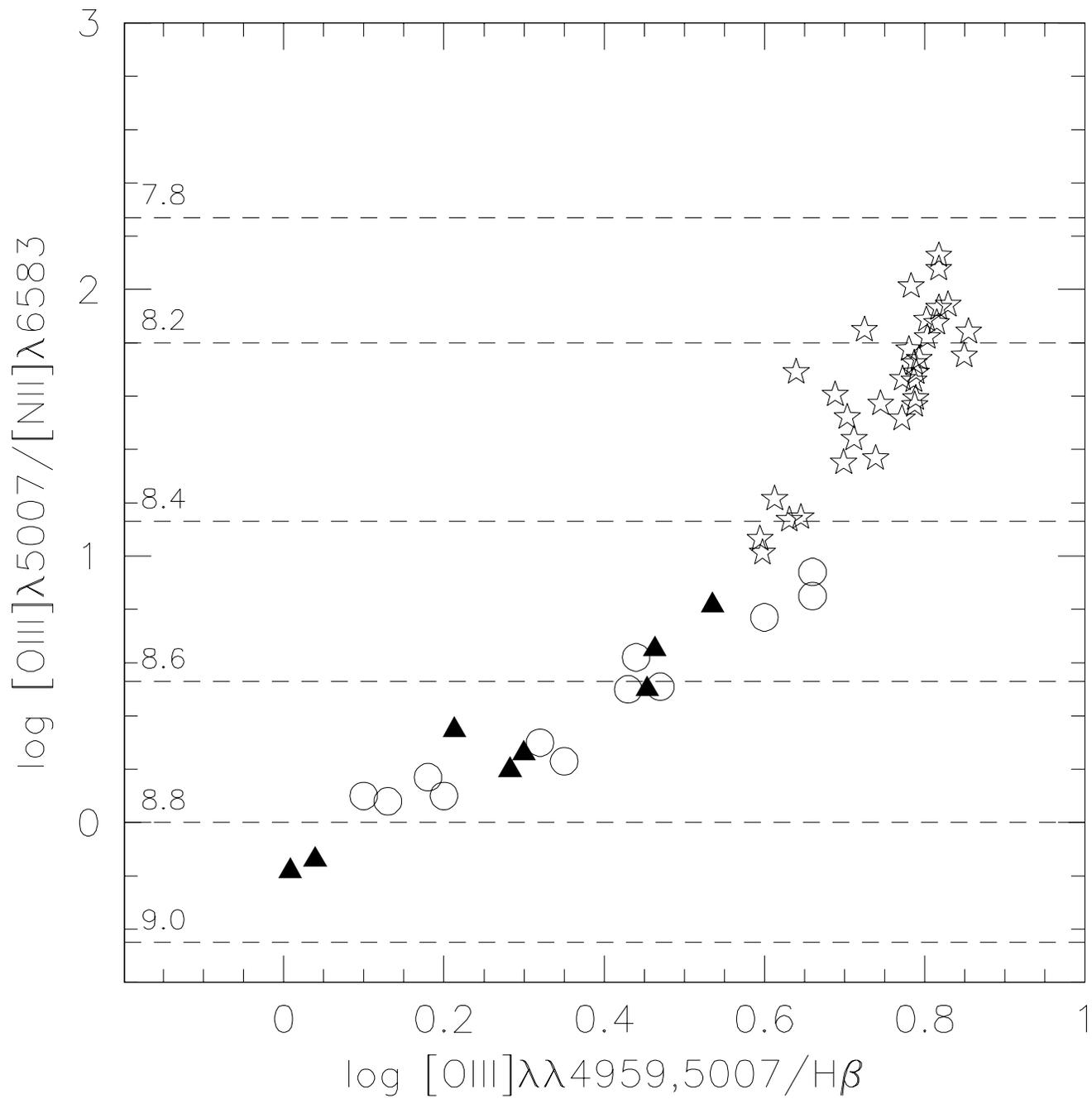
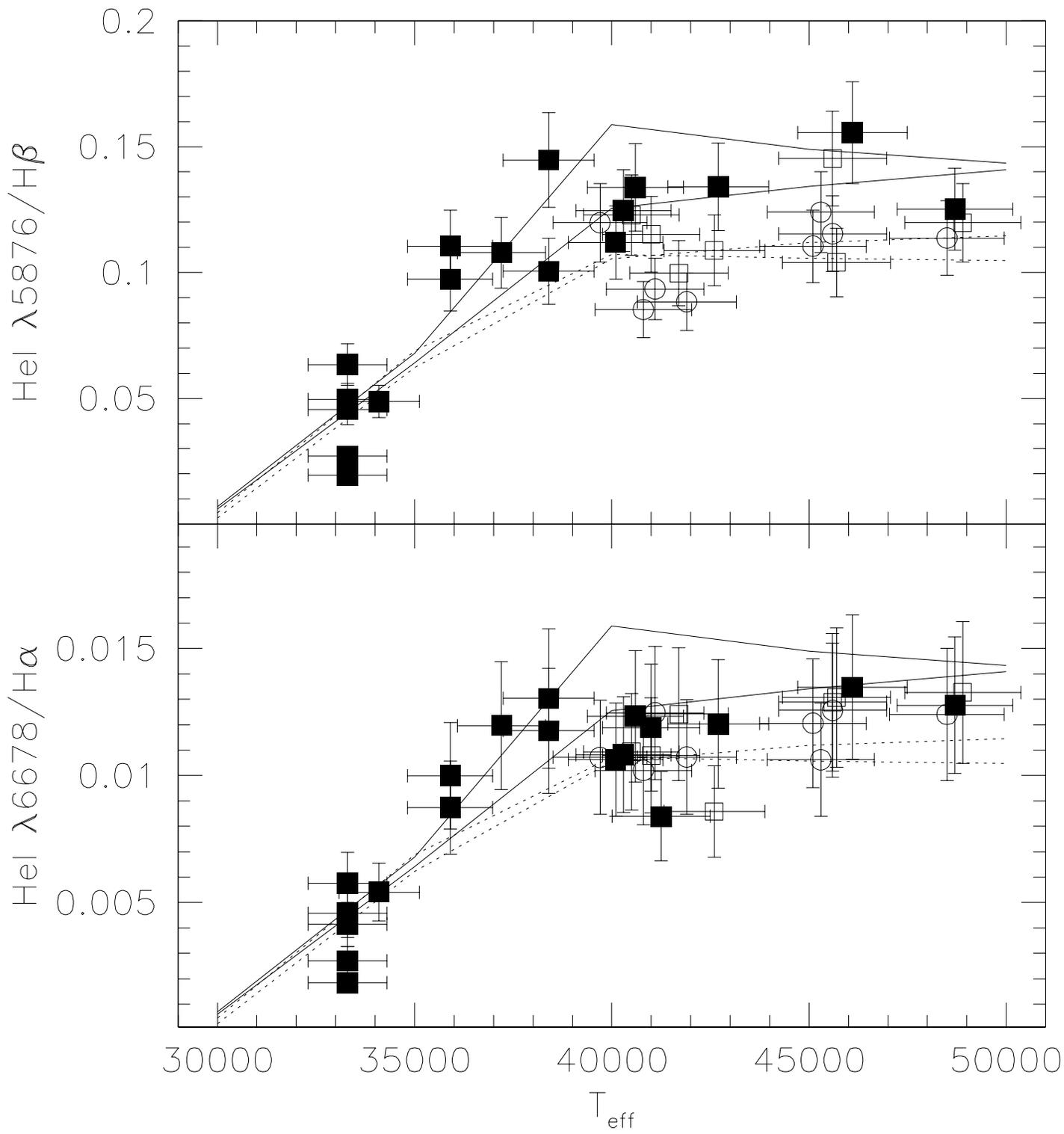
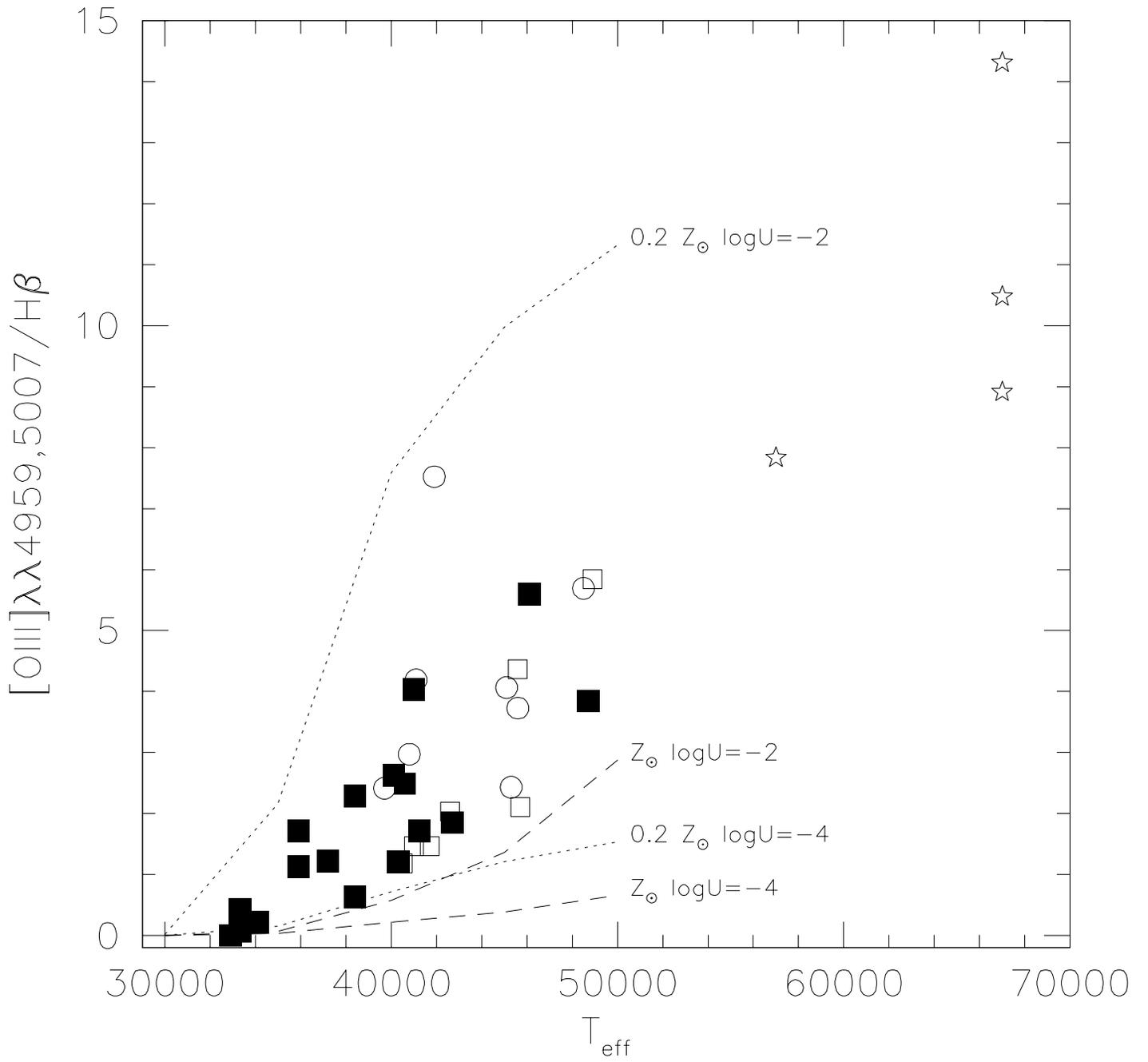
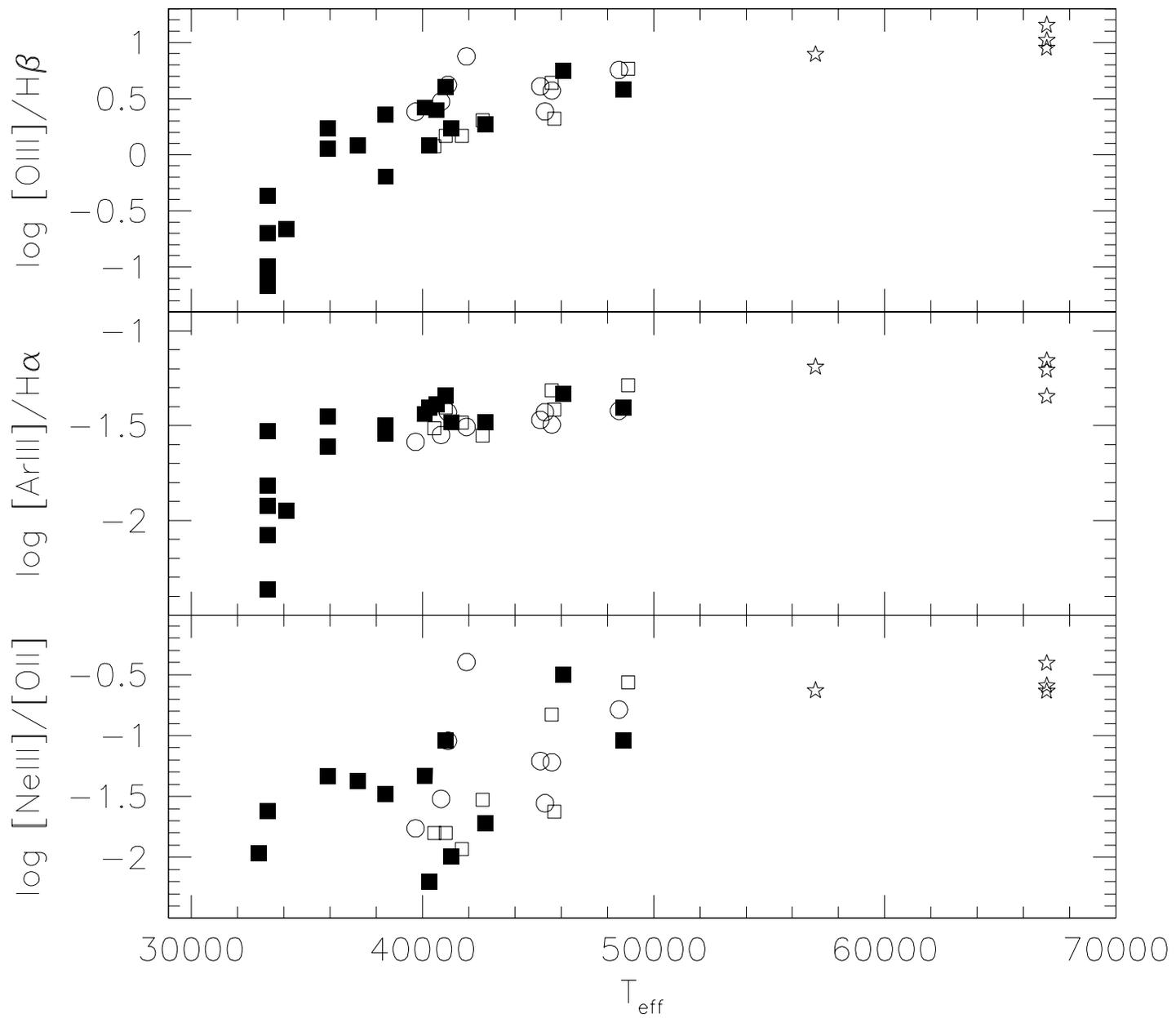
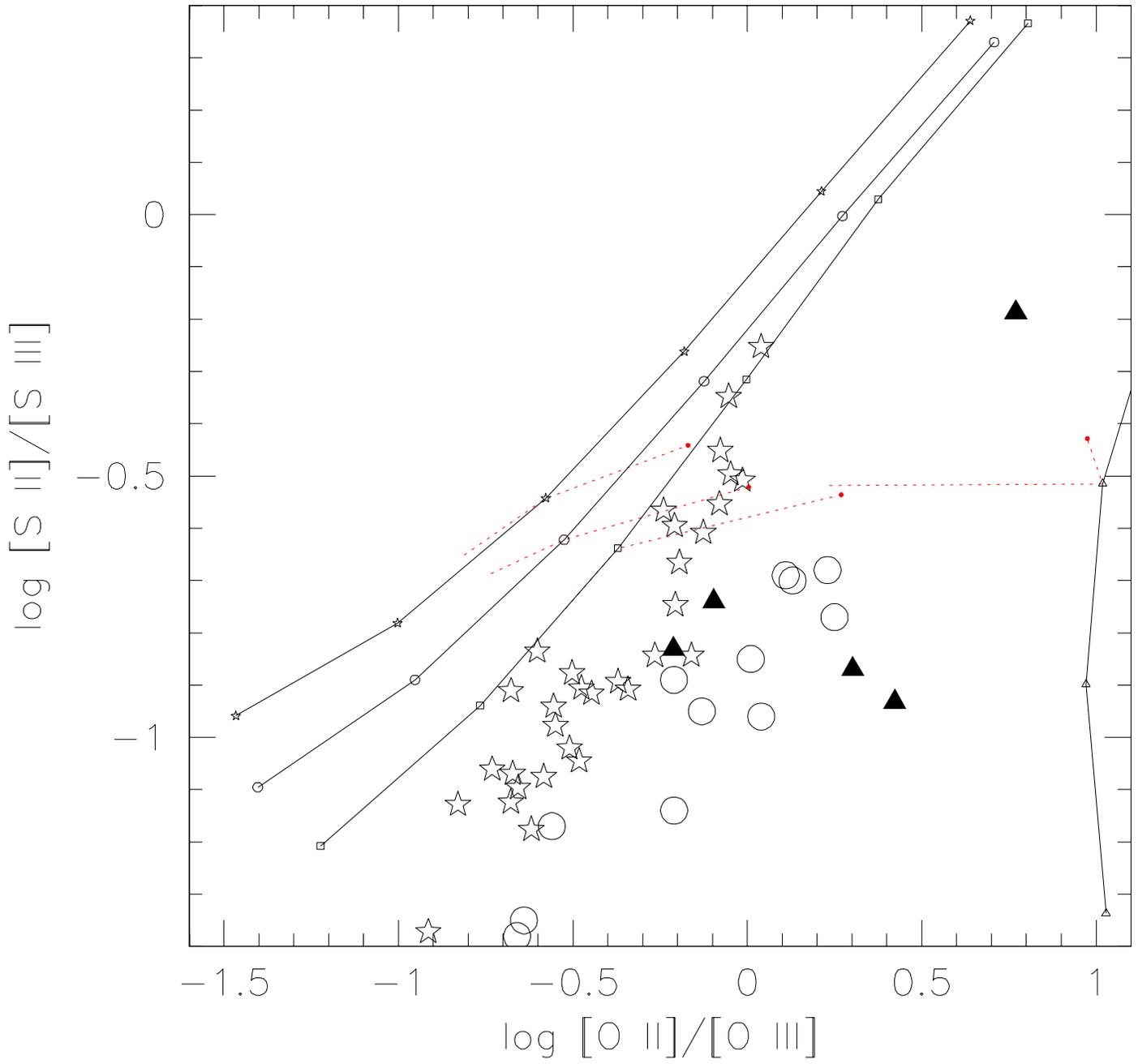
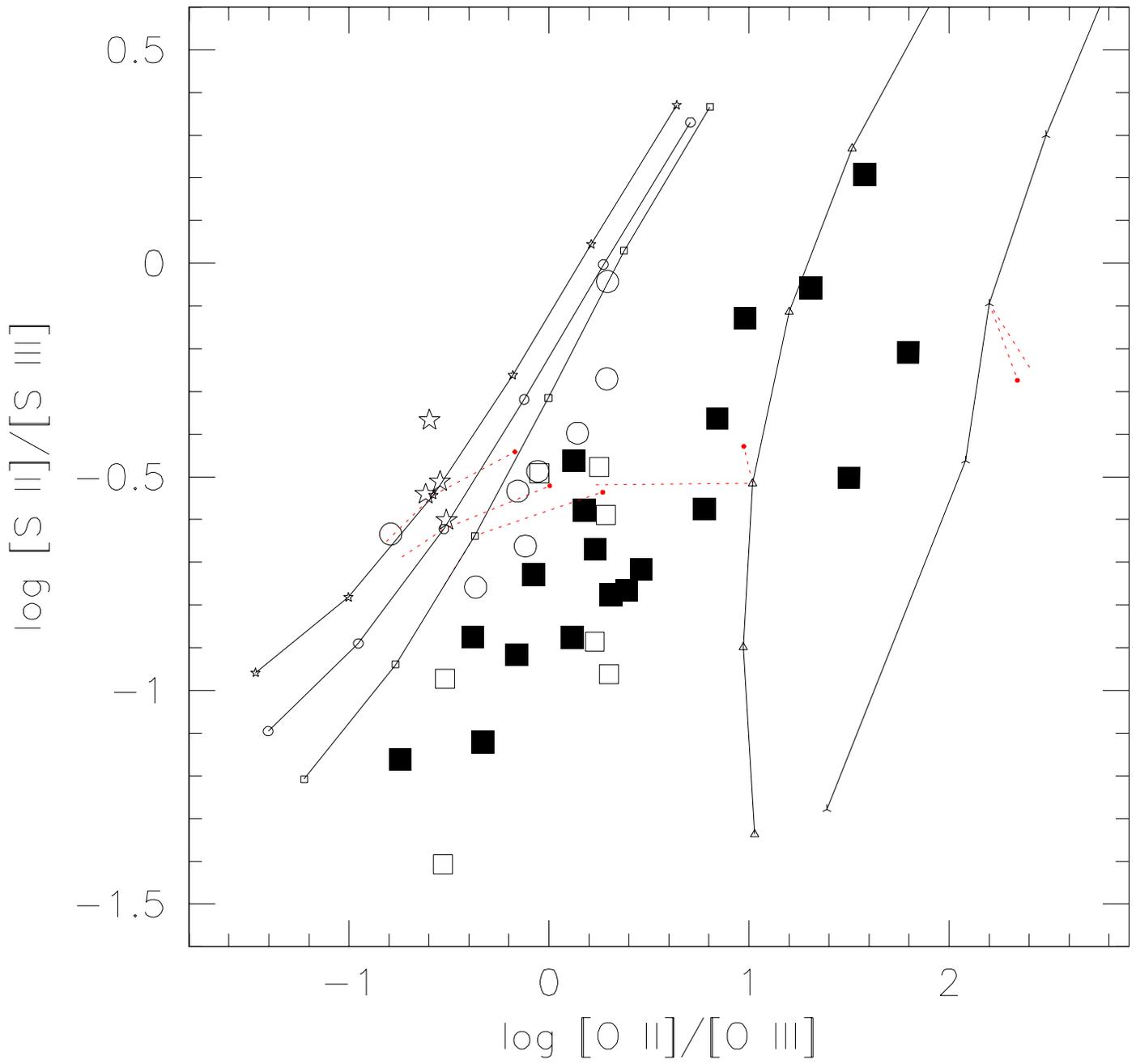
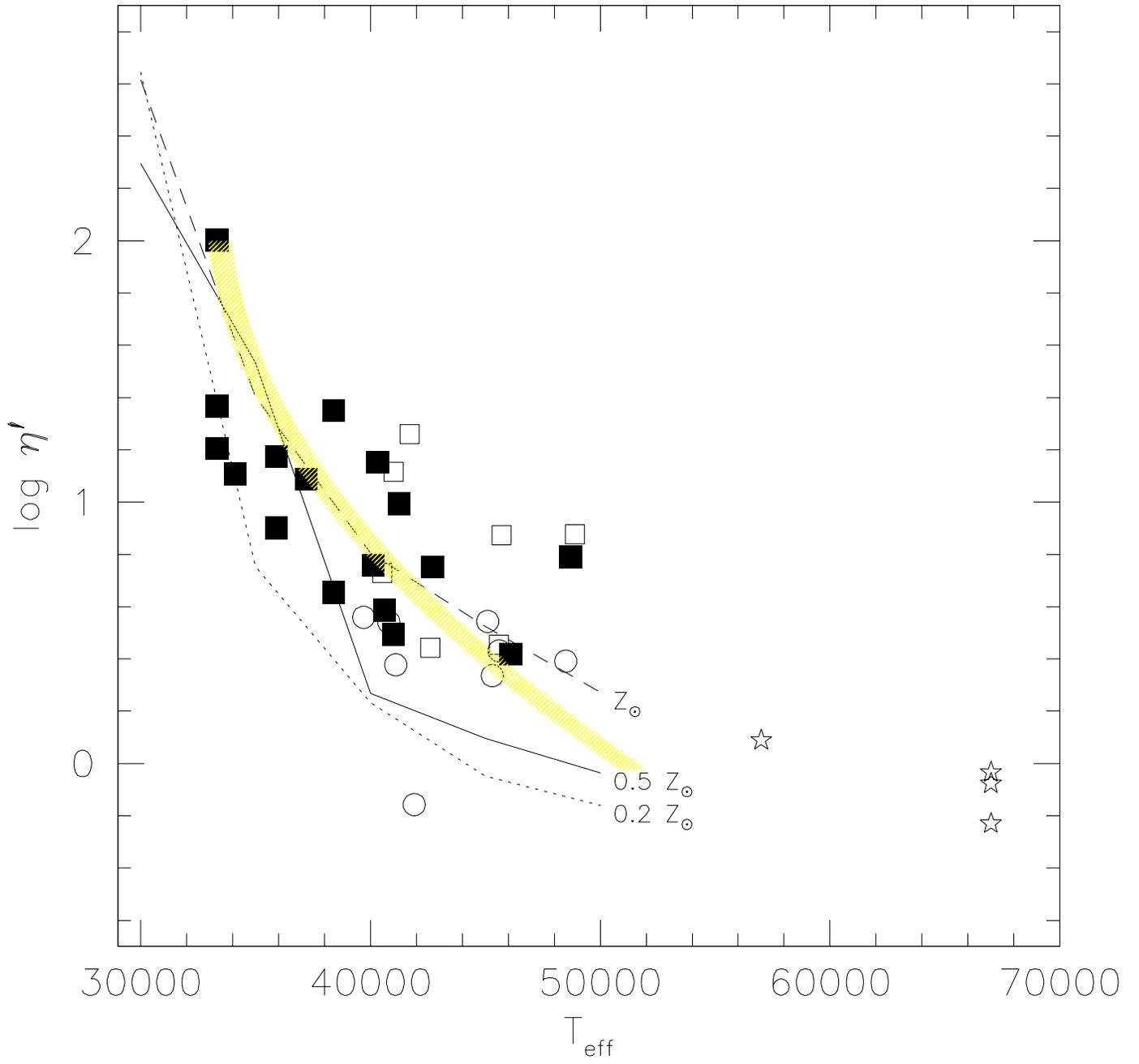
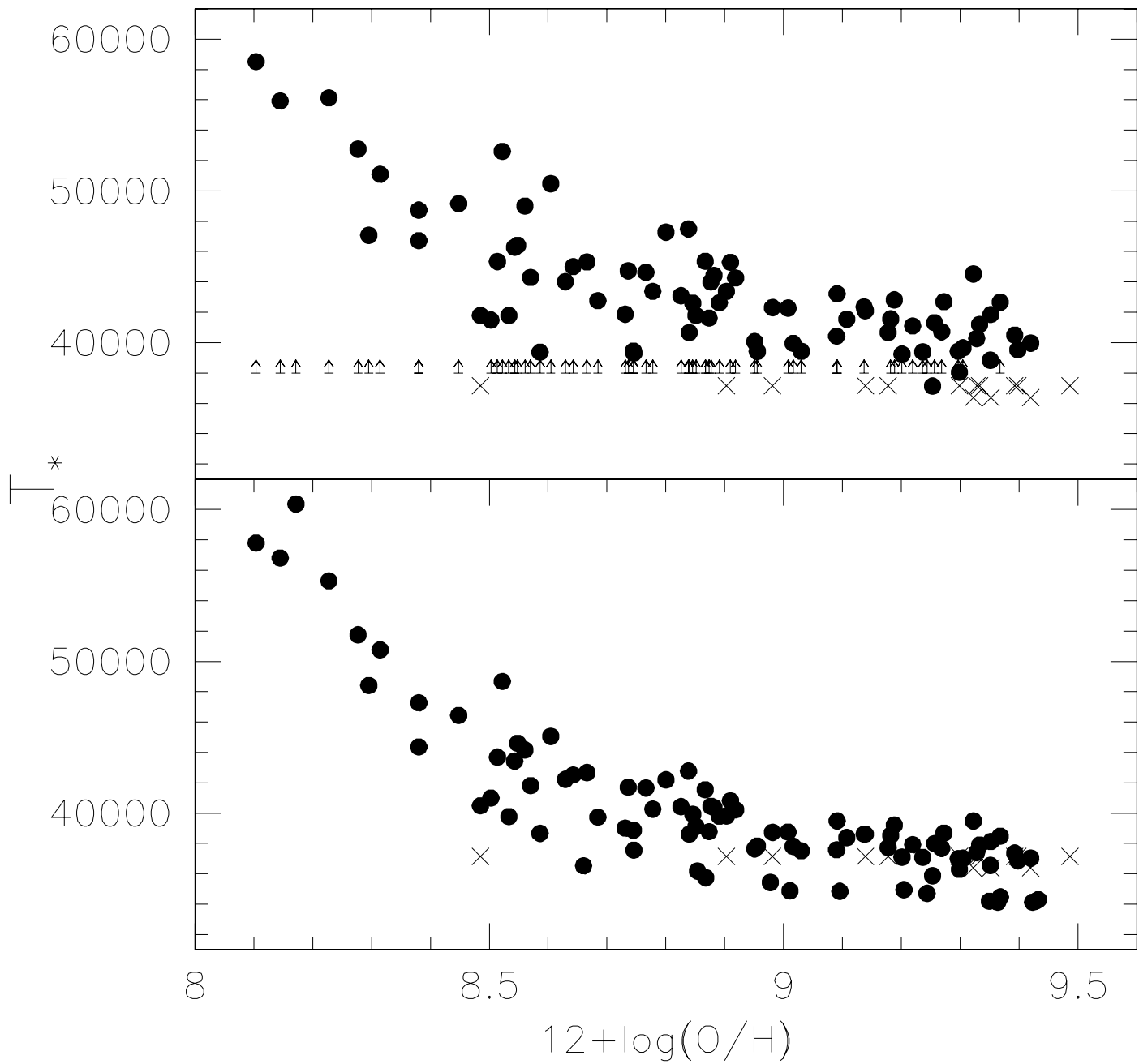
\end{document}